\documentclass[runningheads]{llncs}
\usepackage{graphicx}
\usepackage{comment}
\usepackage{amsmath,amssymb} 
\usepackage{color}

\usepackage[width=122mm,left=12mm,paperwidth=146mm,height=193mm,top=12mm,paperheight=217mm]{geometry}

\usepackage{algorithm}
\usepackage{algorithm,algpseudocode}
\usepackage{xcolor}

\usepackage{subcaption}
\usepackage{algpseudocode}

\usepackage{array}
\newcolumntype{x}[1]{>{\centering\let\newline\\\arraybackslash\hspace{0pt}}p{#1}}

\newcommand\blfootnote[1]{%
  \begingroup
  \renewcommand\thefootnote{}\footnote{#1}%
  \addtocounter{footnote}{-4}%
  \endgroup
}

\begin{document}

\setlength{\belowcaptionskip}{-0.3cm}
\setlength{\abovecaptionskip}{0.2cm}

\pagestyle{headings}
\mainmatter
\def\ECCVSubNumber{s}  
\title{Content Adaptive and Error Propagation Aware Deep Video Compression} 

\titlerunning{Abbreviated paper title}
%
\titlerunning{Content Adaptive and Error Propagation Aware Deep Video Compression}

\author{Guo Lu*\inst{1} \and
Chunlei Cai*\inst{1} \and
Xiaoyun Zhang\inst{1}\and 
Li Chen\inst{1}\and 
Wanli Ouyang\inst{2,3}\and 
Dong~Xu\inst{2} \and
Zhiyong Gao\inst{1}}

\authorrunning{Guo Lu \textit{et al.}}

\institute{Shanghai Jiao Tong University, China\\
\email{\{luguo2014, caichunlei, xiaoyun.zhang, hilichen, zhiyong.gao\}@sjtu.edu.cn}\and
The University of Sydney, Australia\\
\email{\{wanli.ouyang,dong.xu\}@sydney.edu.au}\and
The University of Sydney, SenseTime Computer Vision Research Group}

\maketitle

\begin{abstract}
Recently, learning based video compression methods attract increasing attention. However, the previous works suffer from error propagation due to the accumulation of reconstructed error in inter predictive coding.
Meanwhile, the previous learning based video codecs are also not adaptive to different video contents.
To address these two problems, we propose a content adaptive and error propagation aware video compression system. Specifically, our method employs a joint training strategy by considering the compression performance of multiple consecutive frames instead of a single frame. Based on the learned long-term temporal information, our approach effectively alleviates error propagation in reconstructed frames.
More importantly, instead of using the hand-crafted coding modes in the traditional compression systems, we design an online encoder updating scheme in our system. 
The proposed approach updates the parameters for encoder according to the rate-distortion criterion but keeps the decoder unchanged in the inference stage. 
Therefore, the encoder is adaptive to different video contents and achieves better compression performance by reducing the domain gap between the training and testing datasets.
Our method is simple yet effective and outperforms the state-of-the-art learning based video codecs on benchmark datasets without increasing the model size or decreasing the decoding speed. \blfootnote{*First two authors contributed equally.}
\end{abstract}

\section{Introduction}

With the increasing amount of video content,  it is a huge challenge to store and transmit videos.
In literature, a large number of algorithms \cite{wiegand2003overview,sullivan2012overview} have been proposed to improve the video compression performance. However, all the traditional video compression algorithms \cite{wiegand2003overview,sullivan2012overview} depend on the hand-designed techniques and highly engineered modules without considering the power of end-to-end learning systems.

Recently, a few learning based image and video compression methods \cite{balle2016end,balle2018variational,minnen2018joint,agustsson2017soft,mentzer2018conditional,li2017learning,Wu_2018_ECCV,lu2019dvc} have been proposed.
For example, Lu \textit{et al.} \cite{lu2019dvc} proposed an end-to-end video compression system by replacing all the key components in the traditional video compression methods with neural networks.

\begin{figure}[!t]
    \centering
    \begin{subfigure}[t]{0.45\textwidth}
    \centering
    \includegraphics[width=1\linewidth]{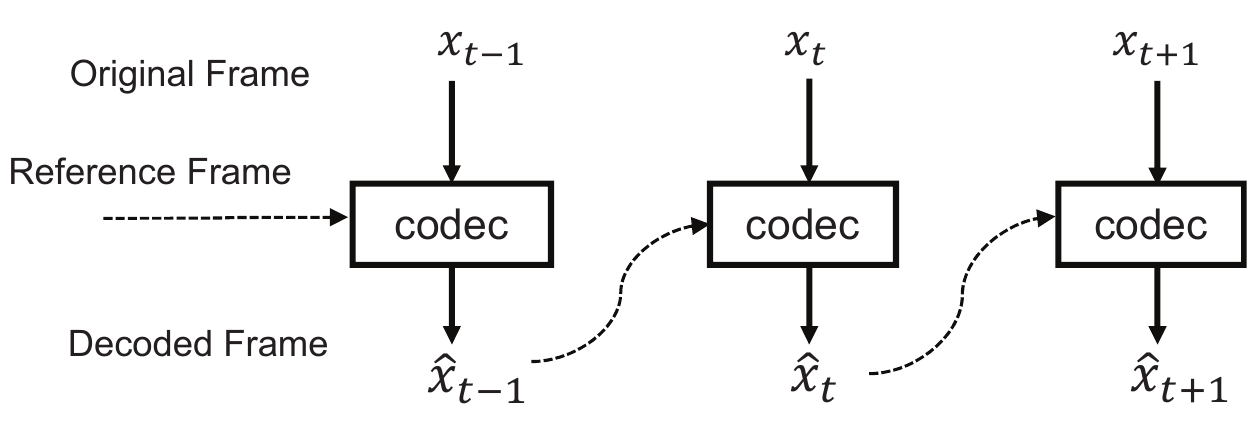}
        \caption{The error propagation issue in the video compression system.} 
    \end{subfigure}
    \begin{subfigure}[t]{0.45\textwidth}
    \centering
    \includegraphics[width=1\linewidth]{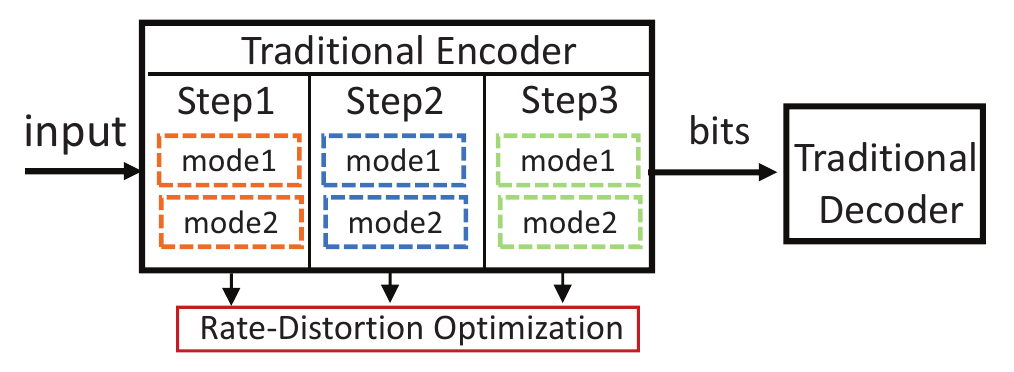}
        \caption{Adaptive mode selection in the traditional video compression system.} 
    \end{subfigure}
    \caption{Two motivations of our proposed method. }
    \label{fig:Motivation}
\end{figure}

However, the current state-of-the-art learning based video compression algorithms \cite{Wu_2018_ECCV,lu2019dvc} still have two drawbacks.
First, the error propagation problem is not considered in the training procedure of learning based video compression systems. As shown in Fig.\ref{fig:Motivation}(a), the previously decoded frame $\hat{x}_{t-1}$ in the coding procedure will be used as the reference frame to compress the current frame $x_t$. Since the video compression is a lossy procedure, the previously decoded frame $\hat{x}_{t-1}$ inevitably has reconstruction error, which will be propagated to the subsequent frames because of the inter-frame predictive coding scheme. As the encoding procedure continues, the error will be accumulated frame by frame and decrease the compression performance significantly.
However, the current approaches \cite{Wu_2018_ECCV,lu2019dvc} train the codecs by only minimizing the distortion between the current frame $x_t$ and the decoded frame $\hat{x}_t$, but ignore the influences of $\hat{x}_t$ on the subsequent encoding process for frame $x_{t+1}$ and so on. 
Therefore, it is critical to building an error propagation aware training strategy for the deep video compression system.

Second, the current learning based encoders \cite{Wu_2018_ECCV,lu2019dvc} are not \textit{adaptive} to different video content as the traditional codecs.  As shown in Fig.\ref{fig:Motivation}(b), the encoder in H.264 \cite{wiegand2003overview} or H.265 \cite{sullivan2012overview} selects different coding modes (\textit{e.g.}, the size of coding unit) for videos with different contents.
In contrast, once the training procedure is finished, the parameters in the learning based encoder are fixed, thus the encoder cannot adapt to different contents in videos and may not be optimal for the current video frame. 
Furthermore, considering the domain gap due to the resolutions or motion magnitudes between training and testing datasets, the learned encoder may achieve inferior performance for videos with some specific contents, such as videos with complex motion scenes.
To achieve content adaptive coding, it is necessary to update the encoder in the inference stage for the learning based video compression system.

In this paper, we propose a content adaptive and error propagation aware deep video compression method. Specifically, to alleviate error accumulation, the video compression system is optimized by minimizing the rate-distortion cost from multiple consecutive frames instead of that from a single frame only.
This joint training strategy exploits the long-term information in the coding procedure, therefore the learning based video codec not only achieves high compression performance for the current frame but also guarantees that the decoded current frame is also useful for the coding procedure of the subsequent frames. Furthermore, we propose an online encoder updating scheme to improve video compression performance.
Instead of using the hand-crafted modes in H.264/H.265,  the parameters of the encoder will be updated based on the rate-distortion objective for \textit{each} video frame.
Our scheme enables the adaption of the encoder according to different video content while keeping the decoder unchanged. 
Experimental results demonstrate the superiority of the proposed method over the traditional codecs.
Our approach is simple yet effective and outperforms the state-of-the-art method \cite{lu2019dvc} without increasing the model size or computational complexity in the decoder side.

The contributions of our work can be summarized as follows,

\begin{enumerate}
\item  An error propagation aware (EPA) training strategy is proposed by considering more temporal information to alleviate error accumulation for the learning based video compression system.

\item We achieve content adaptive video compression in the inference stage by allowing the online update of the video encoder.

\item The proposed method does not increase the model size or computational complexity of the decoder and outperforms the state-of-the-art learning based video codecs.

\end{enumerate}

\vspace{-0.51cm}

\section{Related Work}
\label{sec:relatedwork}

\subsection{Image Compression}

Traditional image compression methods \cite{wallace1992jpeg,skodras2001jpeg,BPG,WebP} use hand-crafted techniques, such as discrete cosine transform (DCT)\cite{ahmed1974discrete} and discrete wavelet transform (DWT) \cite{shensa1992discrete} to reduce the spatial redundancy. 
Recently, learning based image compression approaches attracted increasing attention \cite{toderici2015variable,toderici2017full,balle2016end,balle2018variational,theis2017lossy,agustsson2017soft,li2017learning,rippel2017real,mentzer2018conditional,agustsson2018generative,minnen2018joint,choi2019variable}. 
In \cite{balle2016end}, a CNN based end-to-end image compression framework is proposed by considering both the rate and distortion terms. Furthermore, to obtain the accurate probability model of each symbol, Ball{\'{e}} \textit{et al.} \cite{balle2018variational} estimate the hyperprior for the compressed features and improves the performance of entropy coding.
Since the image compression methods rely on intra prediction, the error propagation reduction issue is not exploited in the existing image compression work.

\vspace{-0.5cm}

\subsection{Video Compression}

The traditional video compression methods \cite{wiegand2003overview,sullivan2012overview} follow the classical block based hybrid coding framework, which uses motion-compensated prediction and transform coding. 
Although each module is well-designed, the traditional video compression systems cannot benefit from the power of deep neural networks.

Recently, more and more end-to-end  frameworks \cite{chen2018learning,Wu_2018_ECCV,lu2019dvc,rippel2018learned,habibian2019video,cheng2019learning,tsai2018learning} were proposed for video compression. 
In \cite{Wu_2018_ECCV}, the video compression task is formulated as frame interpolation, in which the motion information is compressed by using the traditional image compression method \cite{WebP}.
Lu \textit{et al.} \cite{lu2019dvc} proposed a fully end-to-end video compression system. Their approach follows the hybrid coding framework and uses neural networks to implement all components in video compression.
In \cite{rippel2018learned}, an end-to-end video compression framework is proposed and the corresponding motion and residual information are jointly compressed.
Habibian \textit{et al.} \cite{habibian2019video} use a 3D auto-encoder to build a video compression framework and employ an auto-regressive prior as the entropy model.
In \cite{DjelouahICCV2019}, the residual information is computed in the latent space and the proposed framework can directly decode the motion and blending coefficients.
It is worth mentioning that the methods \cite{Wu_2018_ECCV,DjelouahICCV2019,cheng2019learning} are based on frame interpolation and designed for B-frame compression. 

We would like to highlight that the learning based video codecs in these methods \cite{chen2018learning,Wu_2018_ECCV,lu2019dvc,rippel2018learned,habibian2019video,cheng2019learning,tsai2018learning,DjelouahICCV2019} are optimized by minimizing the distortion of a single frame without considering the error propagation problem for videos. More importantly, their encoders are not adaptive to different video content.
Although these methods achieve comparable or even better performance than H.264, we believe that the capability of the existing network architecture is not fully exploited, and the video compression performance can be further improved by using our proposed methods.

\section{Motivations Related to Learning Based Video Compression System}

\begin{figure}[t]
\begin{minipage}[t]{0.45\textwidth}
    \centerline{\includegraphics[width=\linewidth]{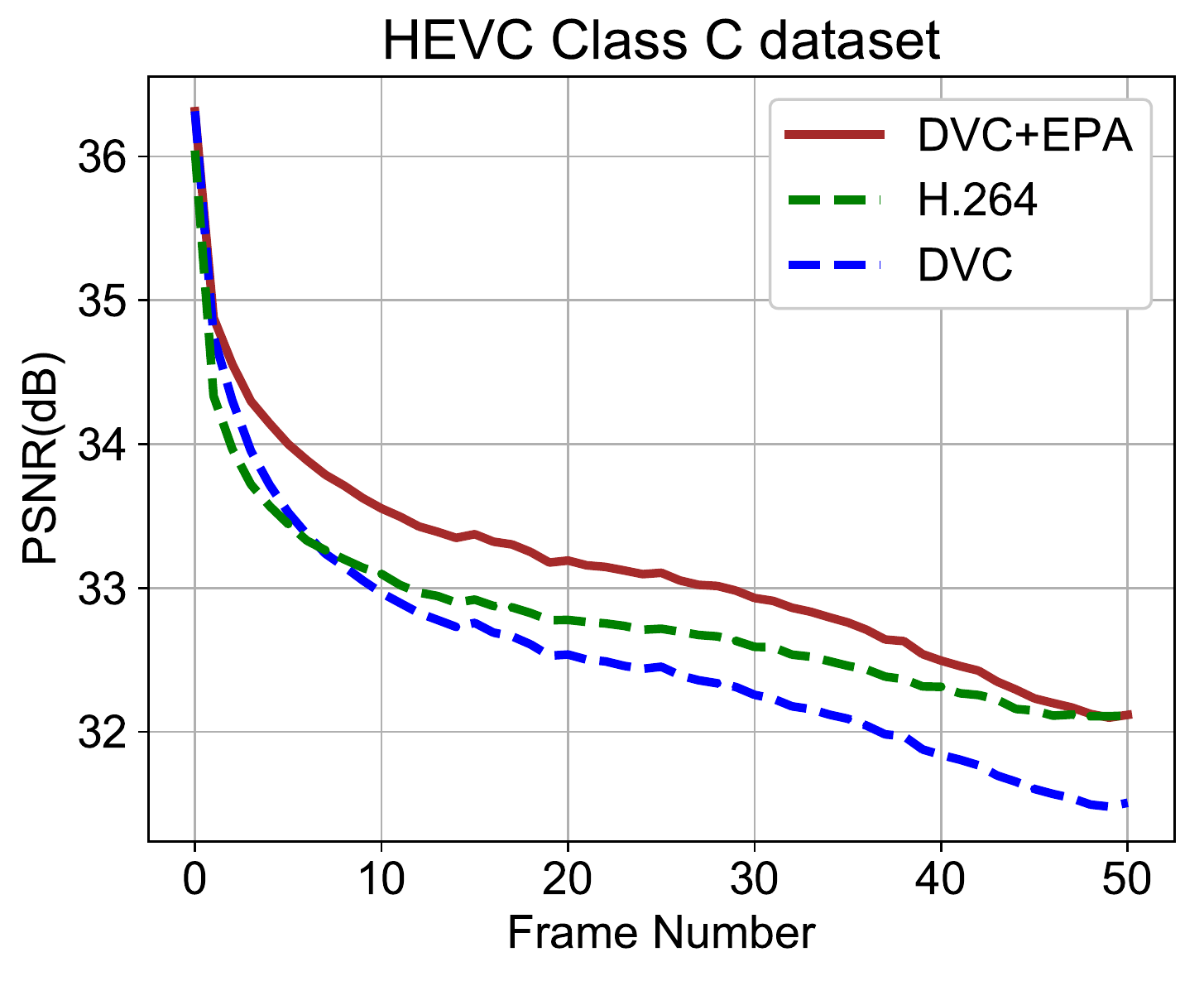}}
    \caption{The PSNR values of reconstructed frames from different algorithms.}
    \label{fig:PSNR_Drop}
\end{minipage}
\begin{minipage}[t]{0.5\textwidth}
    \centerline{\includegraphics[width=\linewidth]{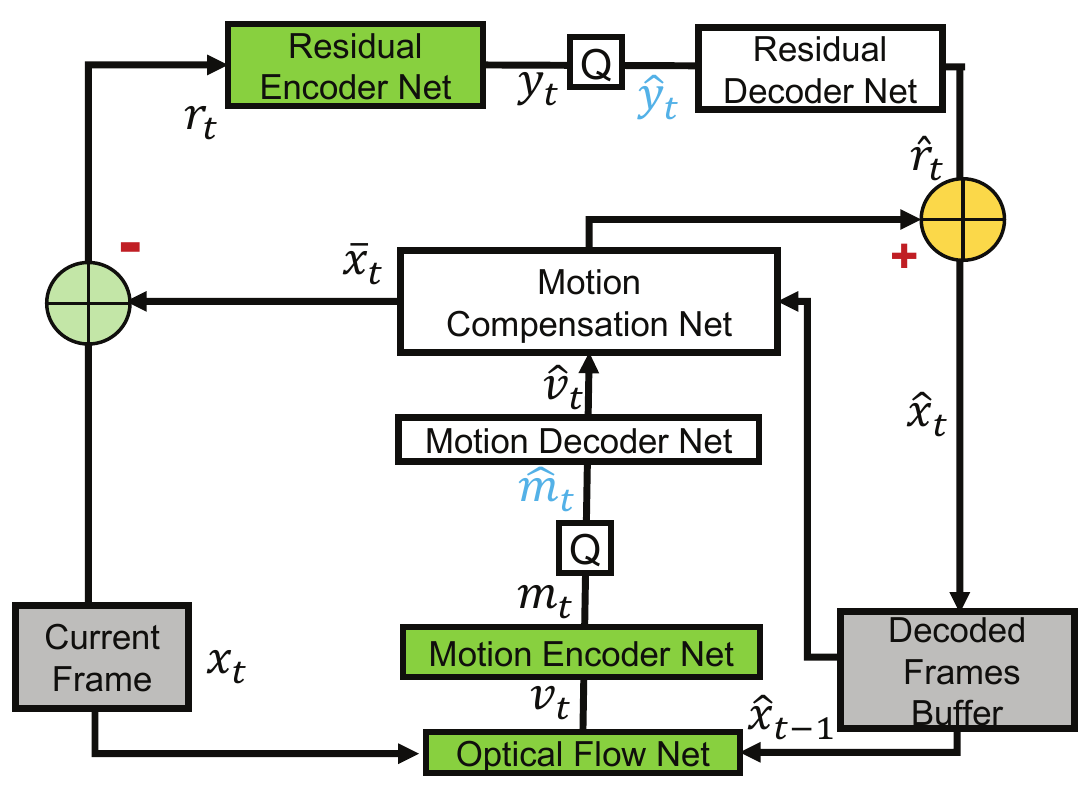}}
    \caption{The architecture of DVC in \cite{lu2019dvc}.}
    \label{fig:OverviewOfDVC}
\end{minipage}
\end{figure}

\subsection{Error Propagation}

Error propagation is a common issue in the video compression systems, mainly due to the inter-prediction.  
In Fig.~\ref{fig:PSNR_Drop}, we provide the PSNRs of the reconstructed frames from the H.264 algorithm \cite{schwarz2007overview} and the learning based video codec DVC \cite{lu2019dvc}.
It is obvious that the PSNR drops when the time step increases.
A possible explanation is that video compression is a lossy procedure and the encoding procedure of the current frame relies on the previous reconstructed frame, which is distorted and thus the error propagates to the subsequent frames.
Let us take the DVC model as an example. The PSNR value of the $5^{th}$ reconstructed frame is 33.52dB,
while the PSNR value of the $6^{th}$ reconstructed frame is 33.37dB(0.15dB drop). 
Furthermore, as the time step increases, the PSNR of the $50^{th}$ reconstructed frame is only 31.50dB.

Although error propagation is inevitable for such a predictive coding framework, it is possible and beneficial to alleviate the error propagation issue and further improve the compression performance (see the curve DVC+EPA in Fig.~\ref{fig:PSNR_Drop}).

\subsection{The Content Adaptive Coding Scheme}

To improve compression performance, the traditional video encoders \cite{wiegand2003overview,sullivan2012overview} use the rate-distortion costs to select the optimal mode for the current frame.
For example, the encoder prefers to use a large block size for homogeneous regions while a small block size is adopted for complex regions. 
To this end, the encoder will calculate the rate-distortion cost for each mode in the coding procedure.
In contrast, the current learning based video compression systems \cite{lu2019dvc,Wu_2018_ECCV} do not employ the content adaptive coding scheme.
In other words, the rate-distortion technique is no longer exploited in the inference stage.
Therefore, the compressed features are not optimal for the current frame.


More importantly, the encoders are optimized by the rate-distortion optimization (RDO) technique in the training dataset, due to the domain gap between the training and testing dataset in terms of resolution or motion magnitudes, the learned encoders may be far from optimal for the testing dataset.
For example, the average motion magnitude between neighboring frames in the training dataset is in the range of $[1,8]$ pixels \cite{xue2017video}. However, the motion in some testing datasets (\textit{e.g.,} the HEVC Class C dataset) is much larger and more complex. The experimental results in \cite{lu2019dvc} also indicate that the compression performance on the HEVC Class C dataset decreases when compared with other datasets.

\section{Proposed Method}

\subsection{Introduction of the DVC framework}

In this paper, we use the framework in \cite{lu2019dvc} as our baseline algorithm to demonstrate the effectiveness of our new approach. In \cite{lu2019dvc}, the deep video compression (DVC) framework follows the classical hybrid coding approach and designs two auto-encoder style networks to compress the motion and residual information, respectively. The architecture of DVC is shown in Fig. \ref{fig:OverviewOfDVC}.
The modules with \textbf{\textcolor[rgb]{0.0,0.5,0.0}{green color}} (\textit{i.e.,} the optical flow net, the motion vector encoder net and the residual encoder net) represent the \textbf{Encoder}.
The other modules (\textit{i.e.}, MV decoder, motion compensation net and residual decoder net) represent the \textbf{Decoder}.
Here, we use $\Phi_E$ and $\Phi_D$ to represent the trainable parameters in the Encoder and Decoder, respectively.
In the inference stage, the parameters in both Encoder and Decoder are fixed. 

The DVC model is optimized by minimizing the following rate-distortion (RD) trade-off,

\begin{equation}
L_t = \lambda D_t + R_t= \lambda d(x_t, \hat{x}_t) + [H(\hat{y}_t) + H(\hat{m}_t)]
\label{eq:rdo}
\end{equation}
$L_t$ is the loss function for the current time step $t$.
$d(\cdot , \cdot)$ is the distortion metric between $x_t$ and $\hat{x}_t$. $\hat{y}_t$ and $\hat{m}_t$ are the compressed latent representations from residual and motion information, respectively. $H(\hat{y}_t)$ and $H(\hat{m}_t)$ are the corresponding number of bits used for compressing these latent representations.
It is noticed that the whole network is optimized to minimize the rate-distortion criterion for the current time step $t$.

However, this scheme ignores two critical \textit{\textbf{dependencies}} for learning based video compression.  
First, the compression system, including the encoder and decoder, ignores the potential influence from reconstruction error of $\hat{x}_t$ to the next frame $x_{t+1}$ in the training procedure and thus leads to the error propagation.
Second, the encoder itself is fixed and not depend on the current frame $x_t$, which deteriorates the compression performance in the inference stage.
In the next sections, we will introduce how to address these two dependencies in video compression.


\begin{figure*}[!t]
\centering
  \includegraphics[width=0.8\linewidth]{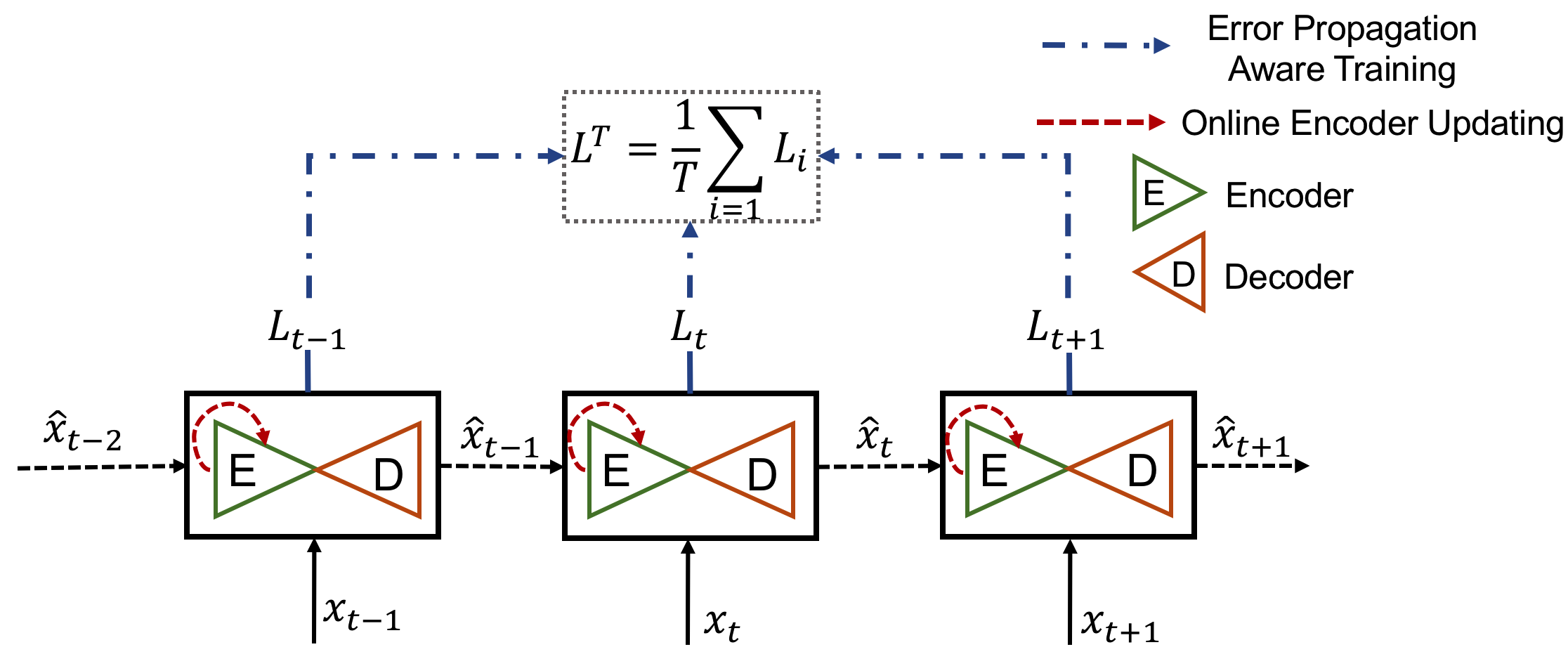}
  \caption{The proposed content adaptive and error propagation aware deep video compression method.}
  \label{fig:proposed}
\end{figure*}

\subsection{The Error Propagation Aware Training Strategy}

To alleviate error accumulation in video compression, we propose an error propagation aware training strategy. 
Specifically, we design a joint training strategy to train the video codec by using the information from different time steps in one video clip and combines all the information to optimize the learned codec for better video compression performance.

The proposed training procedure is shown in Fig.\ref{fig:proposed}. For the current frame $x_t$,  the corresponding reconstructed frame after the encoding and decoding procedure is $\hat{x}_t$. Given $x_t$ and $\hat{x}_t$, we can calculate the RD cost $L_t$.
Then $\hat{x}_t$ will be used as the reference frame in the encoding procedure of $x_{t+1}$, and we obtain the reconstructed frame $\hat{x}_{t+1}$ and the RD cost $L_{t+1}$.  As the coding procedure continues, the reconstructed error will propagate to the following frames. Meanwhile, we also obtain a series of RD costs, which measure the compression performance at the current time step.

Then,  we propose a new objective function by considering the compression performance of both the current frame and the subsequent frames that rely on the current reconstructed frame.
Therefore, the loss function is formulated as follows,
\begin{equation}
\label{eq:TAL}
L^T = \frac{1}{T}\sum_{t=1}^{T} L_t =\frac{1}{T} \sum_{t=1}^{T} \{ \lambda d(x_t, \hat{x}_t) + [H(\hat{y}_t) + H(\hat{m}_t)]\}
\end{equation}
where $T$ is the time interval and  set as 5 in our experiments, $L^T$ represents the error propagation aware loss function.

Therefore, our new training objective will optimize the video codec by employing the objectives from multiple time steps.

As shown in Fig.~\ref{fig:PSNR_Drop}, the video codec DVC with an error propagation aware (DVC+EPA) training strategy significantly reduces error accumulation.
For example, the proposed method has over 0.61dB(32.11dB vs. 31.50dB) improvement over the baseline DVC algorithm \cite{lu2019dvc} for the $50^{th}$ frame, and the gain becomes larger when the time step increases.

\begin{figure}[!ht]
\centering
  \includegraphics[width=0.7\linewidth]{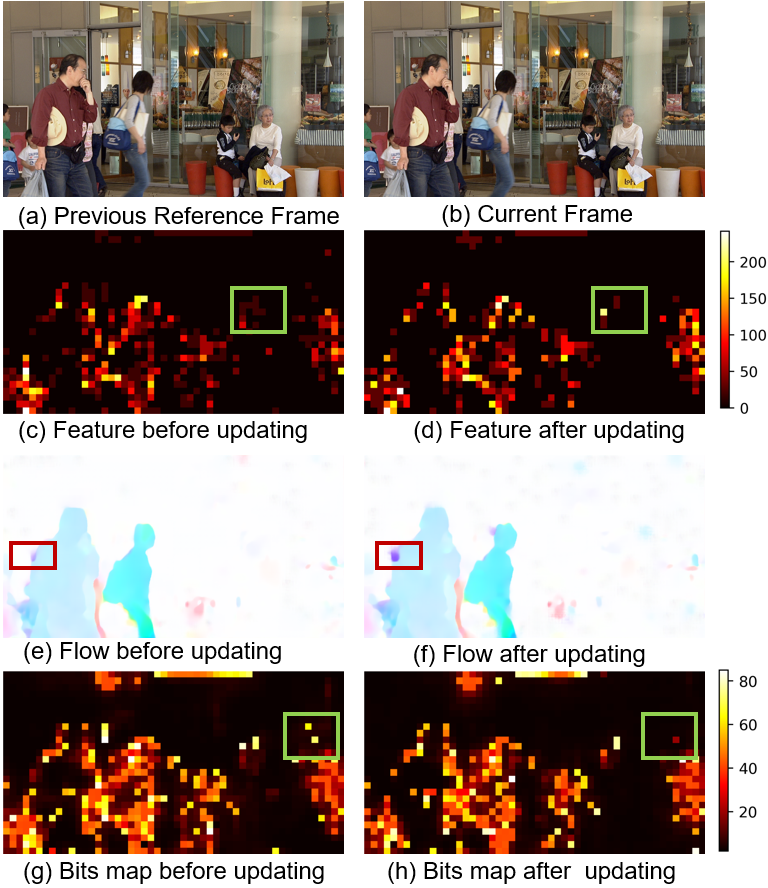}
  \caption{Visual comparison before and after using our online encoder updating scheme.}
  \label{fig:updating}
\end{figure}

\subsection{The Online Encoder Updating Scheme}

To optimize the encoder for each frame and mitigate the domain gap between training and testing data, we propose an online encoder updating scheme in the inference stage.
Our method will update the encoder according to the input image while keeping the decoder unchanged. In other words, we use the training dataset to obtain a general decoder and employ the testing dataset to update the CNN parameters of the encoder.
Based on the training strategy described in the previous section, we can obtain the learned encoder(E) and decoder(D). For the given original frame $x_t$ and the reference frame $\hat{x}_{t-1}$, the objective $L_t$ at the current frame is obtained according to Eq.~\eqref{eq:rdo}.
Then, the parameters $\Phi_D$ in the decoder are fixed while the parameters $\Phi_E$ are updated by minimizing $L_t$.
After several iterations, we obtain the content adaptive encoder, which is optimal for the current frame $x_t$.
Finally, the updated encoder based on testing data and the learned decoder from training data is employed for the actual compression procedure.
The procedure of the proposed online encoder updating algorithm is provided in Algorithm \ref{alg1}.
In our implementation, the maximum iteration number is set to 10. To reduce computational complexity, we will compare  $L_t$ between two consecutive iterations and stop the optimization procedure once the loss becomes stable.

\begin{algorithm}[!t]
\caption{Online Encoder Updating in the Inference Stage}
\label{alg1}
\begin{algorithmic}[1]
\State 
The video encoder \textbf{E} and the trainable parameters $\Phi_E$\\
The video decoder \textbf{D} and the trainable parameters $\Phi_D$\\
The input frame $x_t$ and its reference frame $\hat{x}_{t-1}$\\
$\Delta$ represents the difference between neighbouring iterations
\While{$i <= K$}
\State $\hat{m}_t, \hat{y}_t \Leftarrow \textbf{E}(x_t, \hat{x}_{t-1};\Phi_E^{i-1})$
\State $\hat{x}_t \Leftarrow \textbf{D}(\hat{m}_t, \hat{y}_t;\Phi_D)$
\State $L^i_t= \lambda d(x_t, \hat{x}_t) + [H(\hat{y}_t) + H(\hat{m}_t)]$  
\State $\Phi_E^{i} \Leftarrow  \Phi_E^{i-1} - \alpha \frac{\Delta L_t}{\Delta \Phi_E^{i-1}} $
\If{$ |L^i_t - L^{i-1}_t| < \epsilon$} 
\State $K \Leftarrow i$
\State break
\EndIf
\EndWhile

\State $\hat{m}_t, \hat{y}_t \Leftarrow \textbf{E}(x_t, \hat{x}_{t-1};\Phi_E^{K})$  //Encoding
\State $\hat{x}_t \Leftarrow \textbf{D}(\hat{m}_t, \hat{y}_t;\Phi_D)$ //Decoding

\end{algorithmic}
\end{algorithm}

In contrast to other low-level vision tasks,  the ground-truth frame for video compression is available at the encoder side. As a result, we can update the encoder by using the original frame as long as the decoder remains unchanged.

In Fig. \ref{fig:updating}, we provide the visual results before and after the online updating procedure. 
It is observed that the output feature from the residual encoder (Fig.\ref{fig:updating}(c) and (Fig.\ref{fig:updating}(d)) has changed after the updating procedure which is optimized for the current frames.
More importantly, as shown in Fig.~\ref{fig:updating}(e) and Fig.~\ref{fig:updating}(f), the optical flow map after the updating process contains more details, which is beneficial for accurate prediction.
For example, based on the optical flow in Fig.~\ref{fig:updating}(e), the PSNR of the warped frame is 33.40dB, while the corresponding PSNR of the warped frame is 34.13dB for the updated optical flow in Fig.~\ref{fig:updating}(f).
Furthermore, for the estimated bits map shown in 
Fig.~\ref{fig:updating}(g) and Fig.~\ref{fig:updating}(h), it is observed that the bits map after the updating process allocates fewer bits for the background region.
The experimental results show that the coding bits drop from 0.056bpp to 0.051bpp after the online encoder updating procedure. However, the reconstructed frame has better visual quality after the online updating procedure (36.47dB vs. 36.40dB).

\section{Experiments}

\subsection{Experimental Setup}

\textbf{Datasets}
In the training stage, we use the Vimeo-90k dataset\cite{xue2017video}.  Vimeo-90k is a widely used dataset for low-level vision tasks \cite{wang2019edvr,Lu_2018_ECCV}.
It is also used in the recent learning based video compression tasks \cite{lu2019dvc,DjelouahICCV2019}. 
To evaluate the compression performance of different methods, we employ the following datasets,

\textit{\textbf{HEVC Common Test Sequences}} \cite{sullivan2012overview} are the most popular test sequences for evaluating the video compression performance \cite{sullivan2012overview}. The contents in common test sequences are diversified and challenging. We use Class~B($1920 \times 1080$), Class~C($832 \times 480$) and Class~D($352 \times 288$) in our experiments. 

\textit{\textbf{Video Trace Library(VTL) dataset}} \cite{VTL} contains lots of raw YUV sequences used for the low-level computer vision tasks. Following the setting in \cite{DjelouahICCV2019}, we use 20 video sequences with the resolution of 352 $\times$ 288 in our experiments, and the maximum length of the video clips is set to 300 for all sequences. 

\textit{\textbf{Ultra Video Group(UVG) dataset}} \cite{UVG}  is a high frame rate(120fps) video dataset, in which the motion between neighbouring frames is small. Following the setting in \cite{Wu_2018_ECCV,lu2019dvc,habibian2019video}, we use the video sequences with the resolution of $1920 \times 1080$ in our experiments. 

\textit{\textbf{MCL-JVC dataset}} \cite{MCL} consists of 24 videos with the resolution of 1920 $\times$ 1080. This dataset is widely used for video quality assessment. For a fair comparison with \cite{DjelouahICCV2019}, we also include this dataset in our experiments.


\begin{figure*}[!h]
  \centering
  \begin{minipage}{0.32\textwidth}
    \centerline{\includegraphics[width=\linewidth]{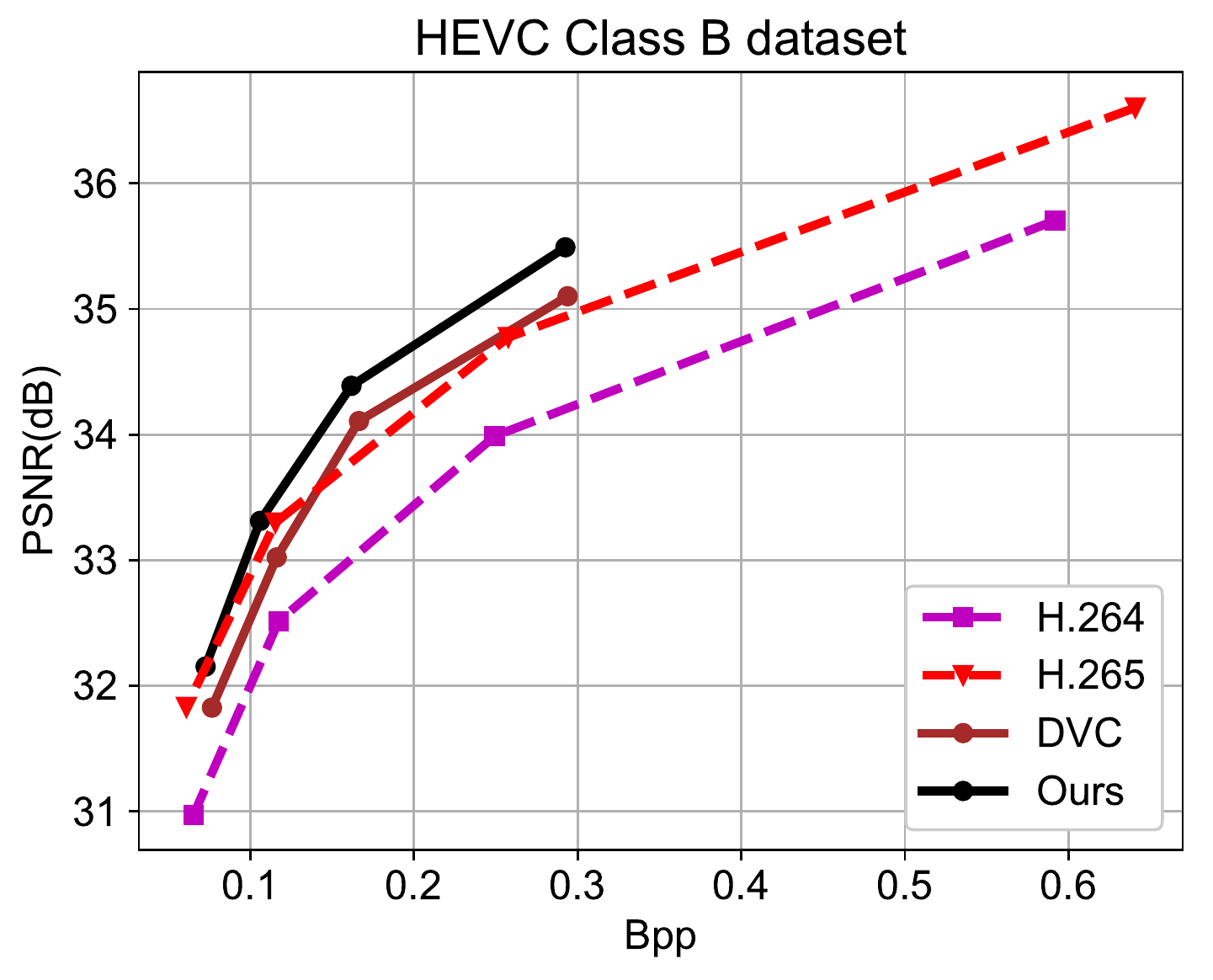}}
  \end{minipage}
    \hspace{\fill}
  \begin{minipage}{0.32\textwidth}
    \centerline{\includegraphics[width=\linewidth]{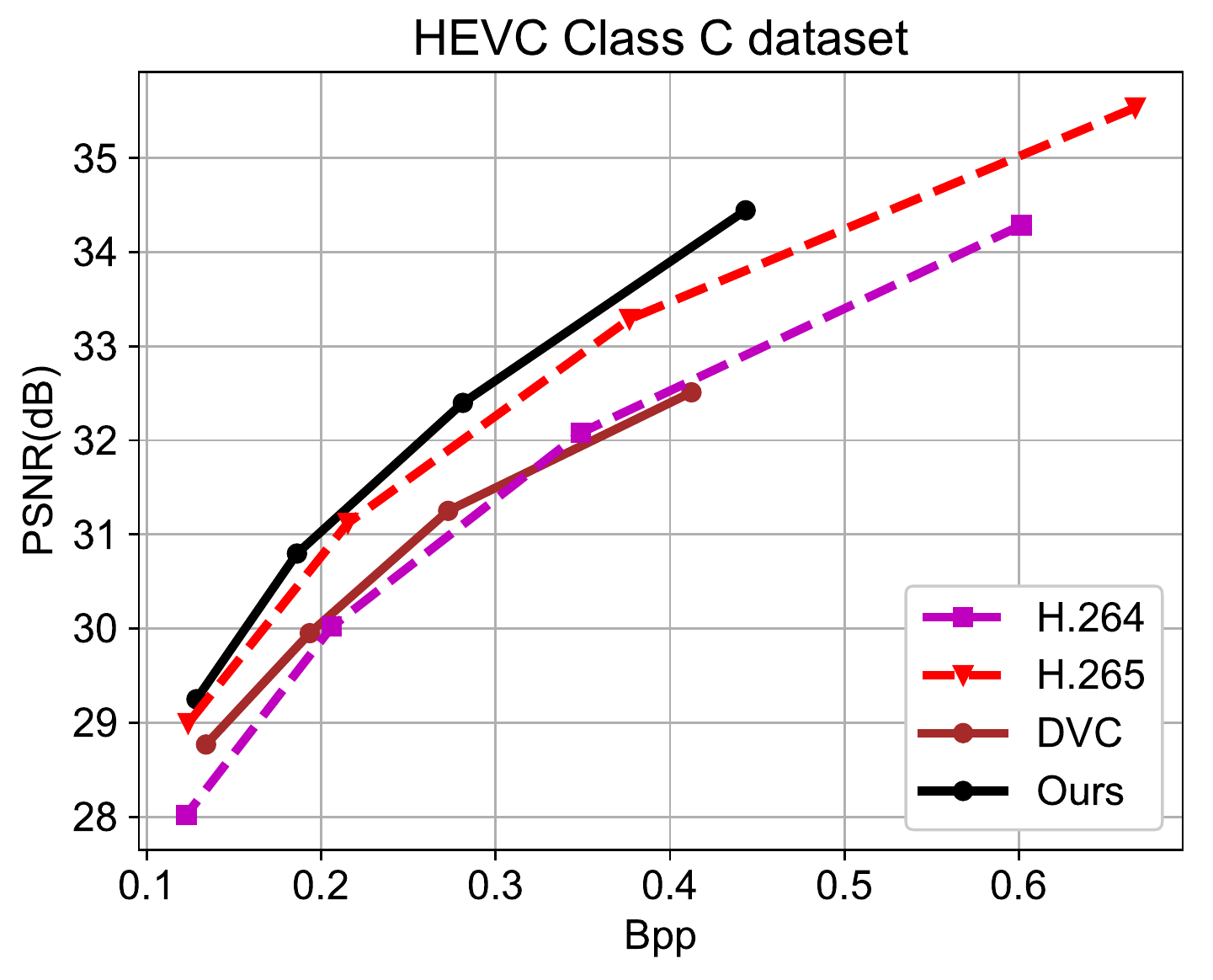}}
  \end{minipage}
  \hspace{\fill}
    \begin{minipage}{0.32\textwidth}
    \centerline{\includegraphics[width=\linewidth]{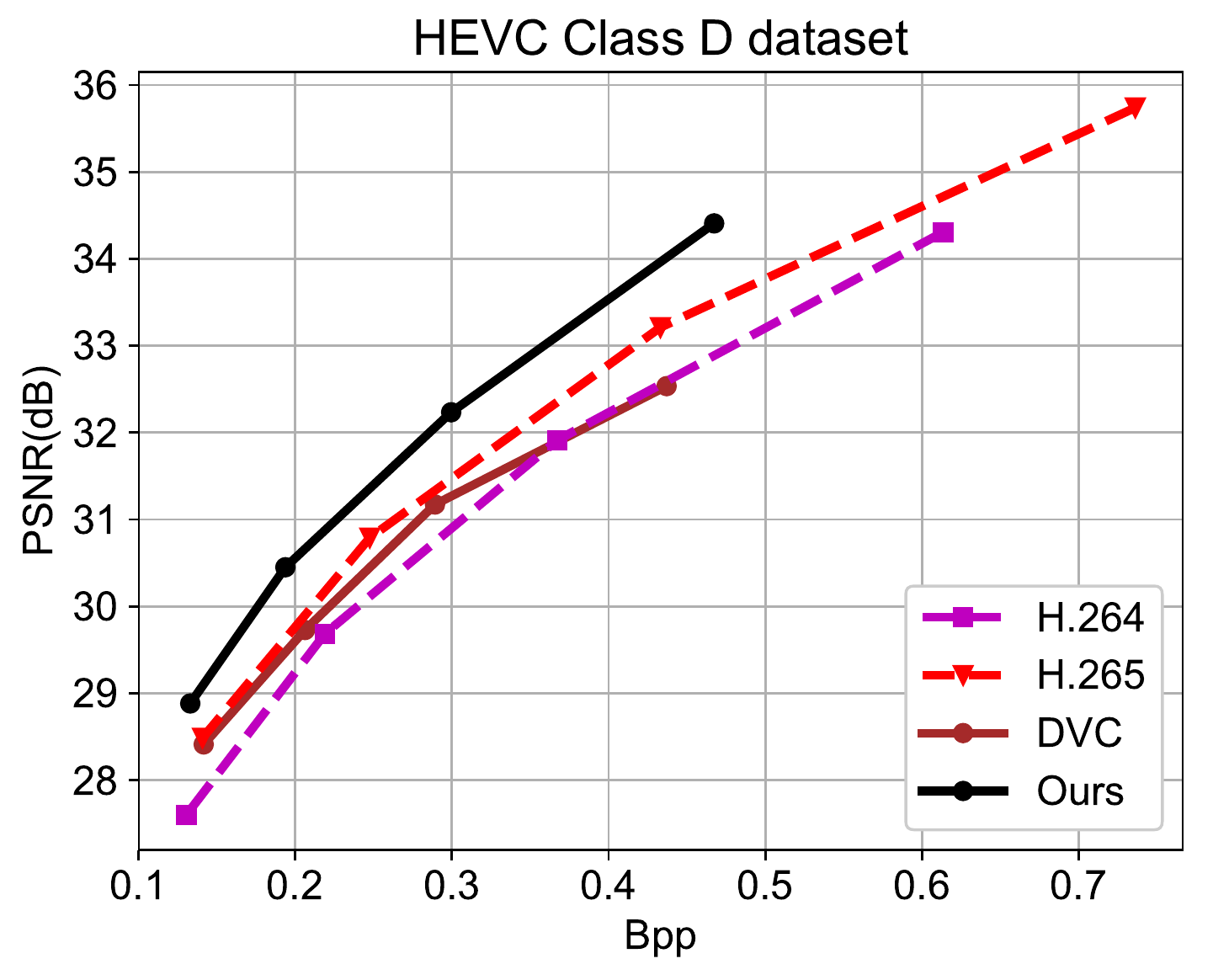}}
  \end{minipage}
  
  \begin{minipage}{0.32\textwidth}
    \centerline{\includegraphics[width=\linewidth]{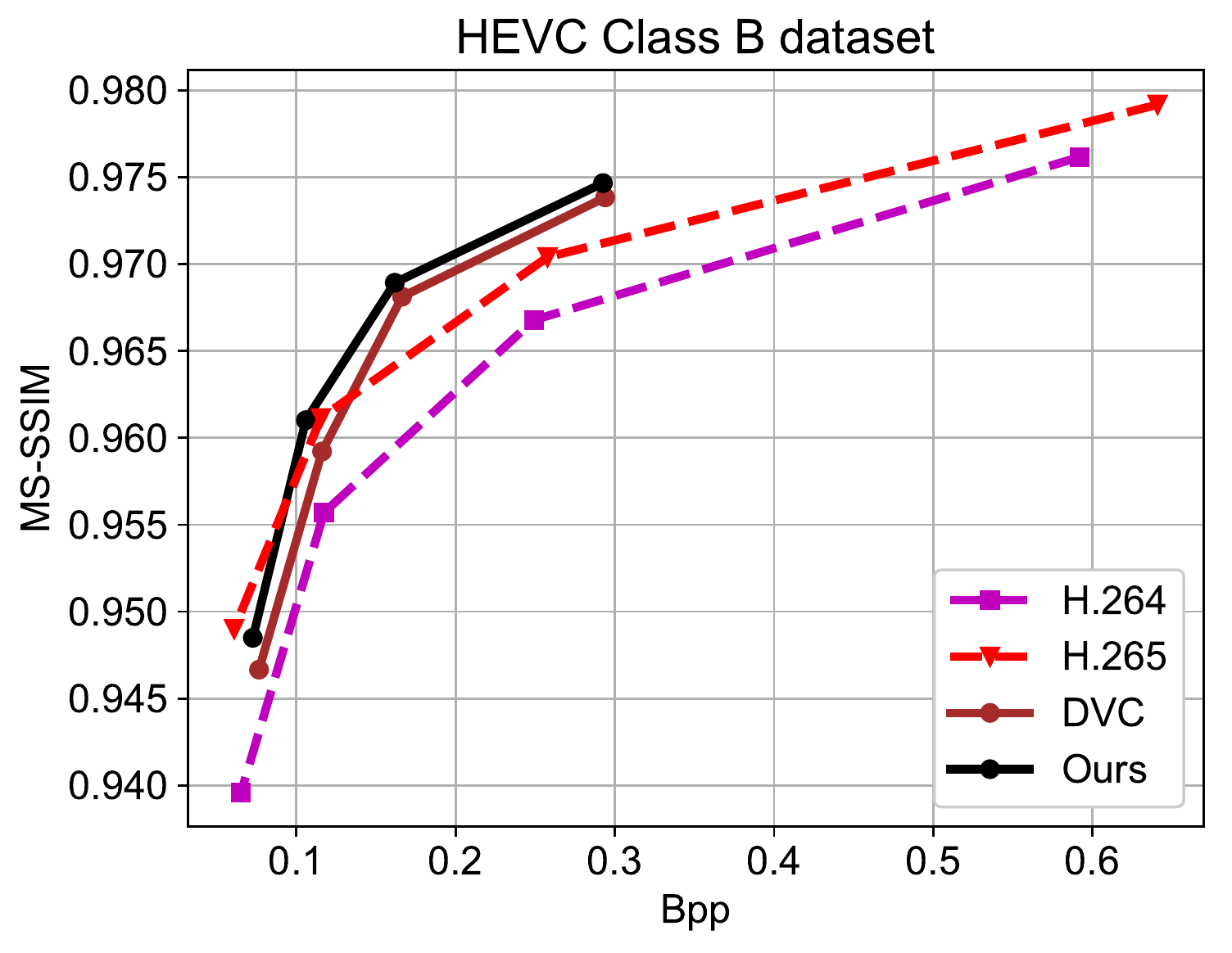}}
  \end{minipage}
    \hspace{\fill}
  \begin{minipage}{0.32\textwidth}
    \centerline{\includegraphics[width=\linewidth]{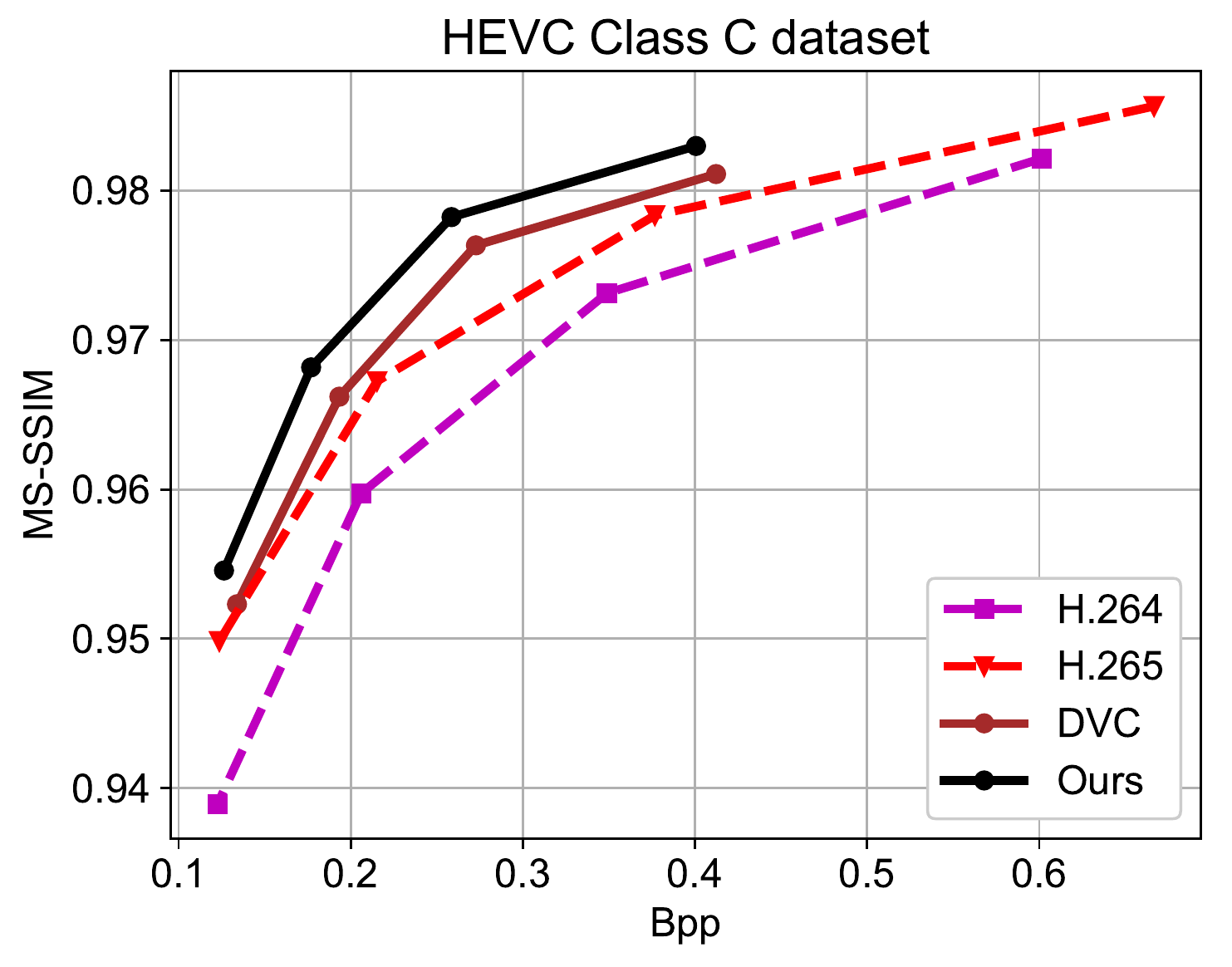}}
  \end{minipage}
  \hspace{\fill}
    \begin{minipage}{0.32\textwidth}
    \centerline{\includegraphics[width=\linewidth]{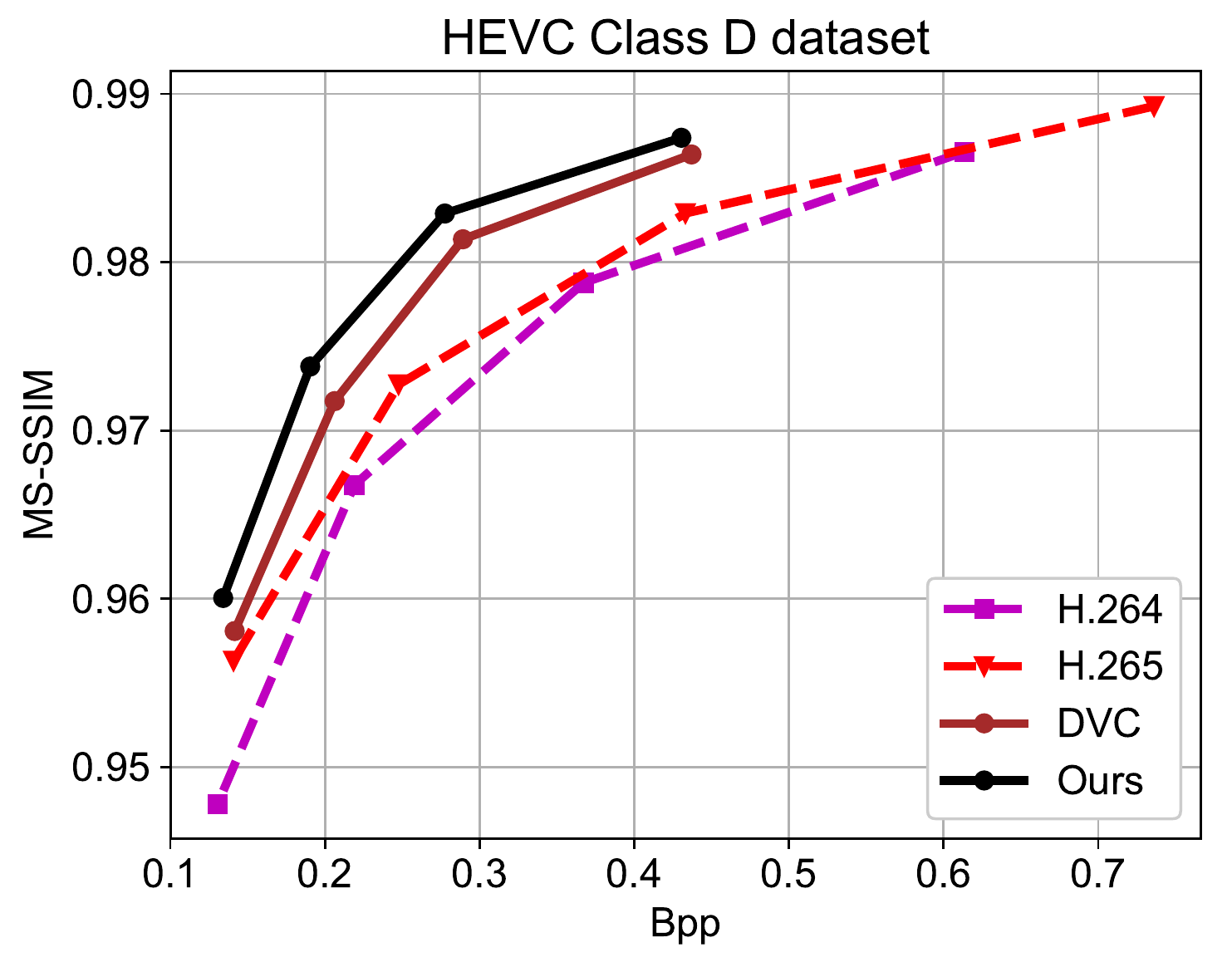}}
  \end{minipage}

  \caption{Comparison between our proposed method with the learning based video codec in \cite{lu2019dvc}, H.264 \cite{wiegand2003overview} and H.265 \cite{sullivan2012overview}.}

  \label{fig:mainresults}
\end{figure*}

\begin{figure*}
  \begin{minipage}{0.32\textwidth}
    \centerline{\includegraphics[width=\linewidth]{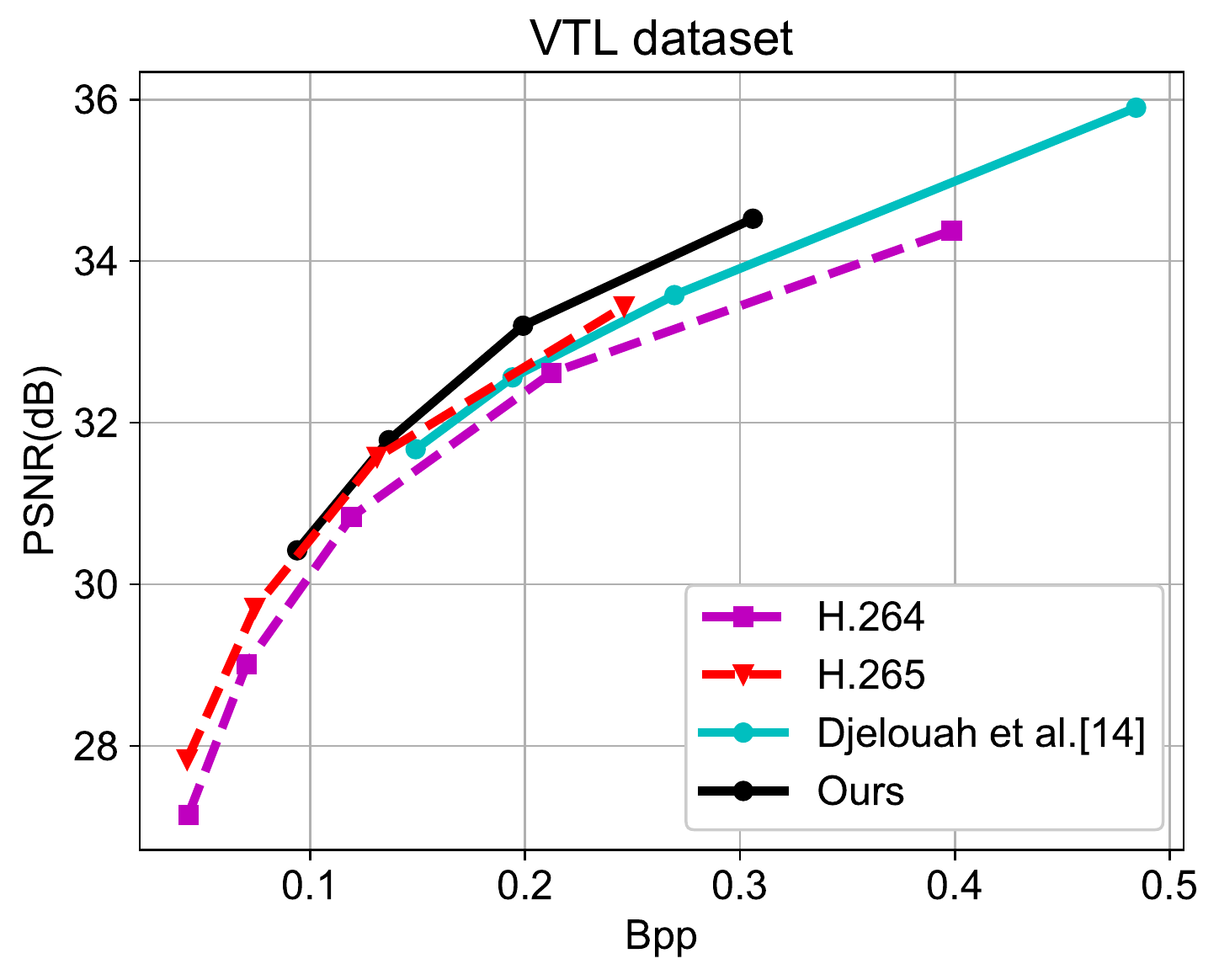}}
    \caption{PSNR performance evaluation on the VTL dataset}
      \label{fig:VTL}
  \end{minipage}
    \hspace{\fill}
  \begin{minipage}{0.32\textwidth}
    \centerline{\includegraphics[width=\linewidth]{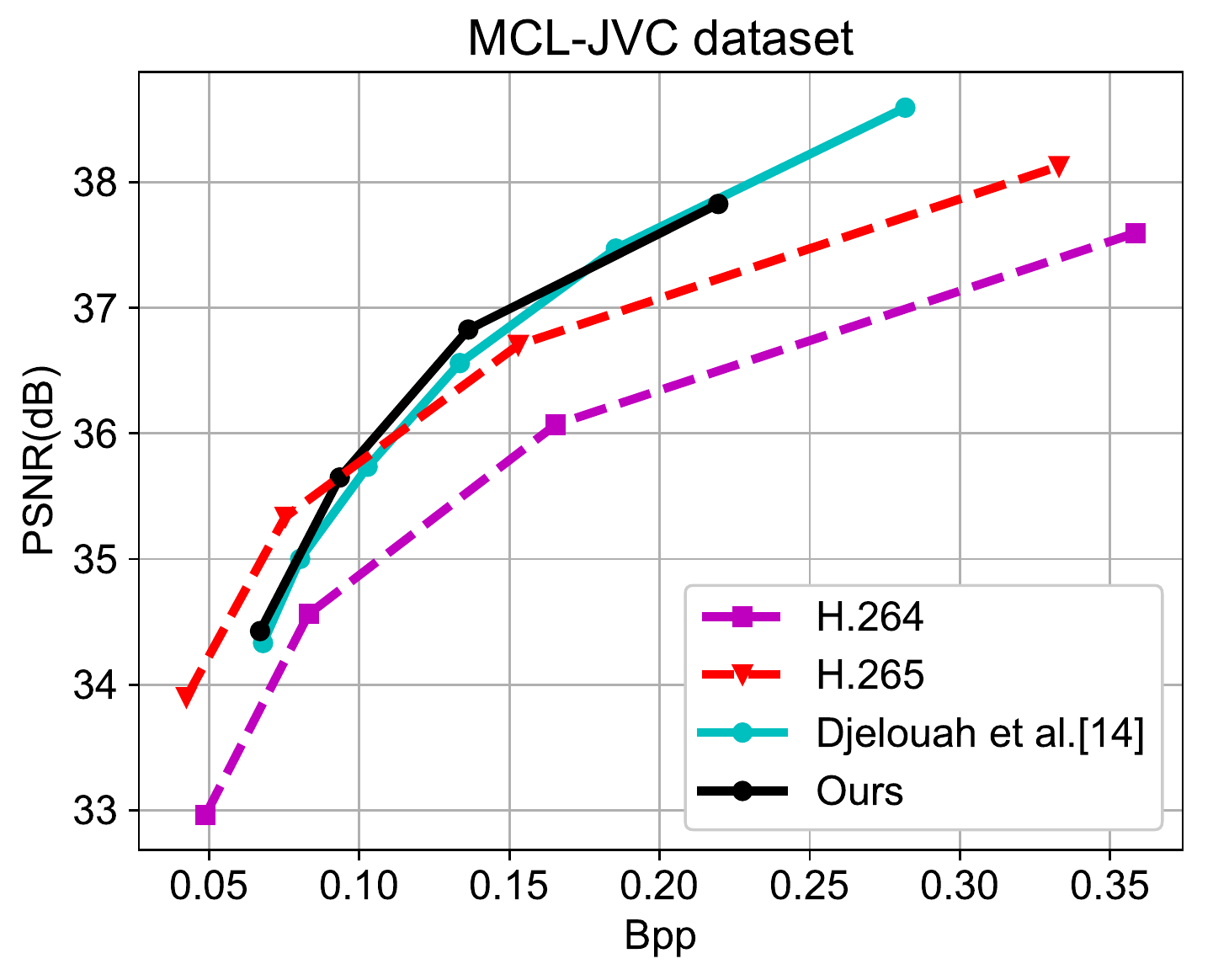}}
    \caption{PSNR performance evaluation on the MCL-JVC dataset}
     \label{fig:MCL}
  \end{minipage}
  \hspace{\fill}
    \begin{minipage}{0.32\textwidth}
    \centerline{\includegraphics[width=\linewidth]{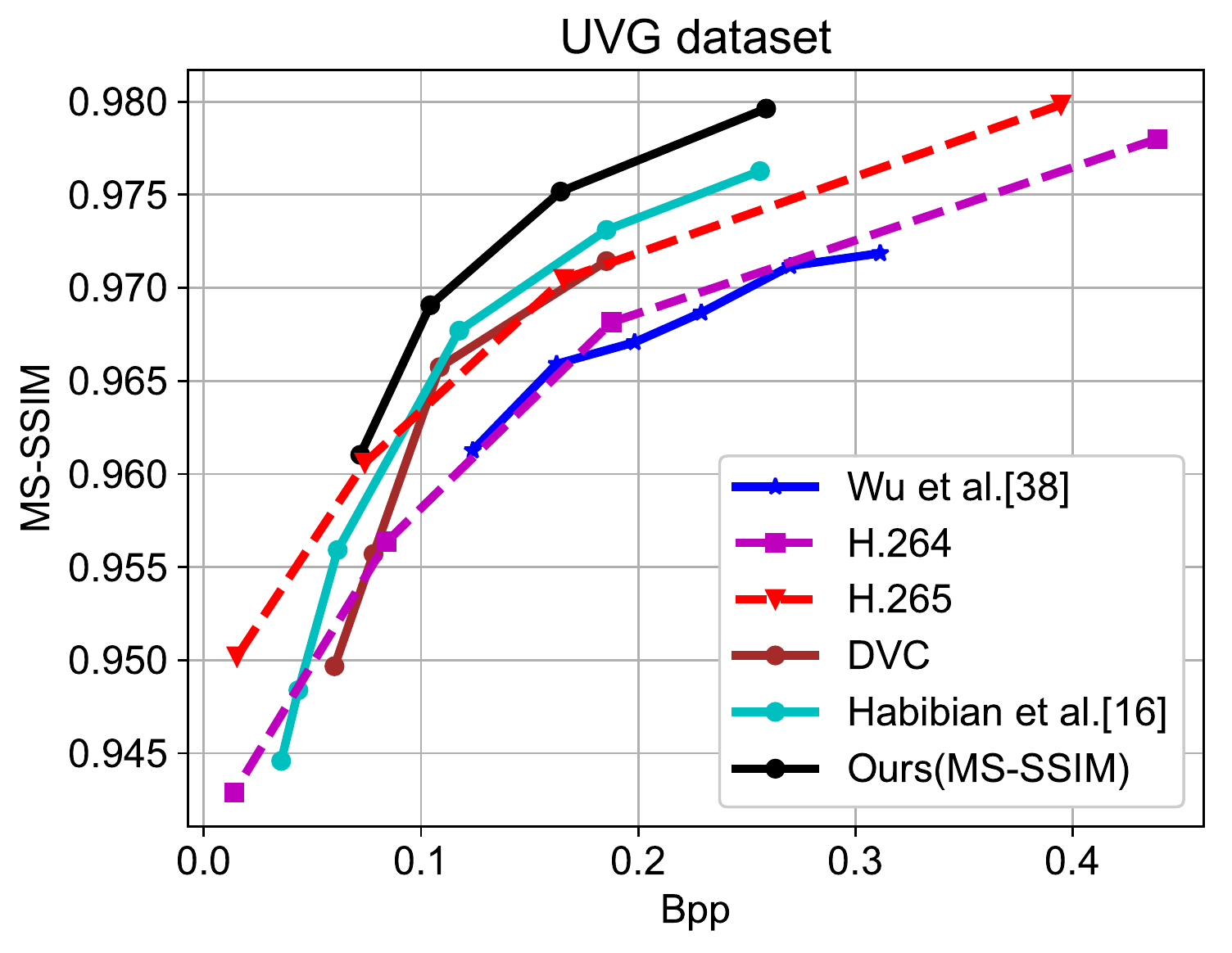}}
    \caption{MS-SSIM performance evaluation on the UVG dataset}
         \label{fig:UVG}
  \end{minipage}
\end{figure*}

\textbf{Implementation details.}
We train four models with different $\lambda$ values (256, 512, 1024, 2048) in Eq.~\eqref{eq:rdo}.
To generate the I-frame/key-frame for video compression, we use the learning based image compression method in \cite{balle2018variational}, in which the corresponding $\lambda$ in the image codec are empirically set to 1024, 2048, 4096 and 12000, respectively.

In our implementation, we use DVC \cite{lu2019dvc} as the baseline method. In the training stage, the whole network is first optimized by using the loss in Eq.\eqref{eq:rdo}, then is fine-tuned based on the error propagation aware loss in Eq.\eqref{eq:TAL}. The corresponding batch sizes are set to 4 and 1, respectively. The resolution of the training images is $256 \times 256$. We use Adam optimizer \cite{kingma2014adam}  and the initial learning rate is set as $1e-4$.
In the inference stage, the encoder is also optimized by using Adam optimizer \cite{kingma2014adam} to achieve content adaptive encoding.
The proposed method is implemented based on Tensorflow. It takes about 5 days to train the whole network by using two GTX 1080Ti GPUs.

In our experiments, we use the PSNR and MS-SSIM \cite{wang2003multi} to measure the distortion between the original frame and the reconstructed frame.
The bits per pixel(bpp) represents the coding bits in the compression procedure.

\subsection{Comparison with the state-of-the-art methods}

\textbf{Evaluation Setting.}
To make a fair comparison with the state-of-the-art learning based video compression methods and the traditional video codecs H.264/H.265, we follow the existing evaluation protocols in \cite{Wu_2018_ECCV,lu2019dvc,DjelouahICCV2019} to perform extensive experiments. Specifically, all the existing learning based methods \cite{Wu_2018_ECCV,lu2019dvc,DjelouahICCV2019} use  \textit{\textbf{fixed GoP setting}}. For example, the GoP size for the UVG dataset is set to 12 while the corresponding GoP size is set to 10 for the HEVC Common Test Sequences \cite{Wu_2018_ECCV,lu2019dvc}. And the corresponding GoP size for H.265/H.264 in these papers is also fixed to 12 or 10. We follow the same settings and provide the experimental results in Fig.\ref{fig:mainresults}, Fig.\ref{fig:VTL}, Fig.\ref{fig:MCL} and Fig.\ref{fig:UVG}.

\begin{figure*}[!t]
  \begin{minipage}{0.32\textwidth}
    \centerline{\includegraphics[width=\linewidth]{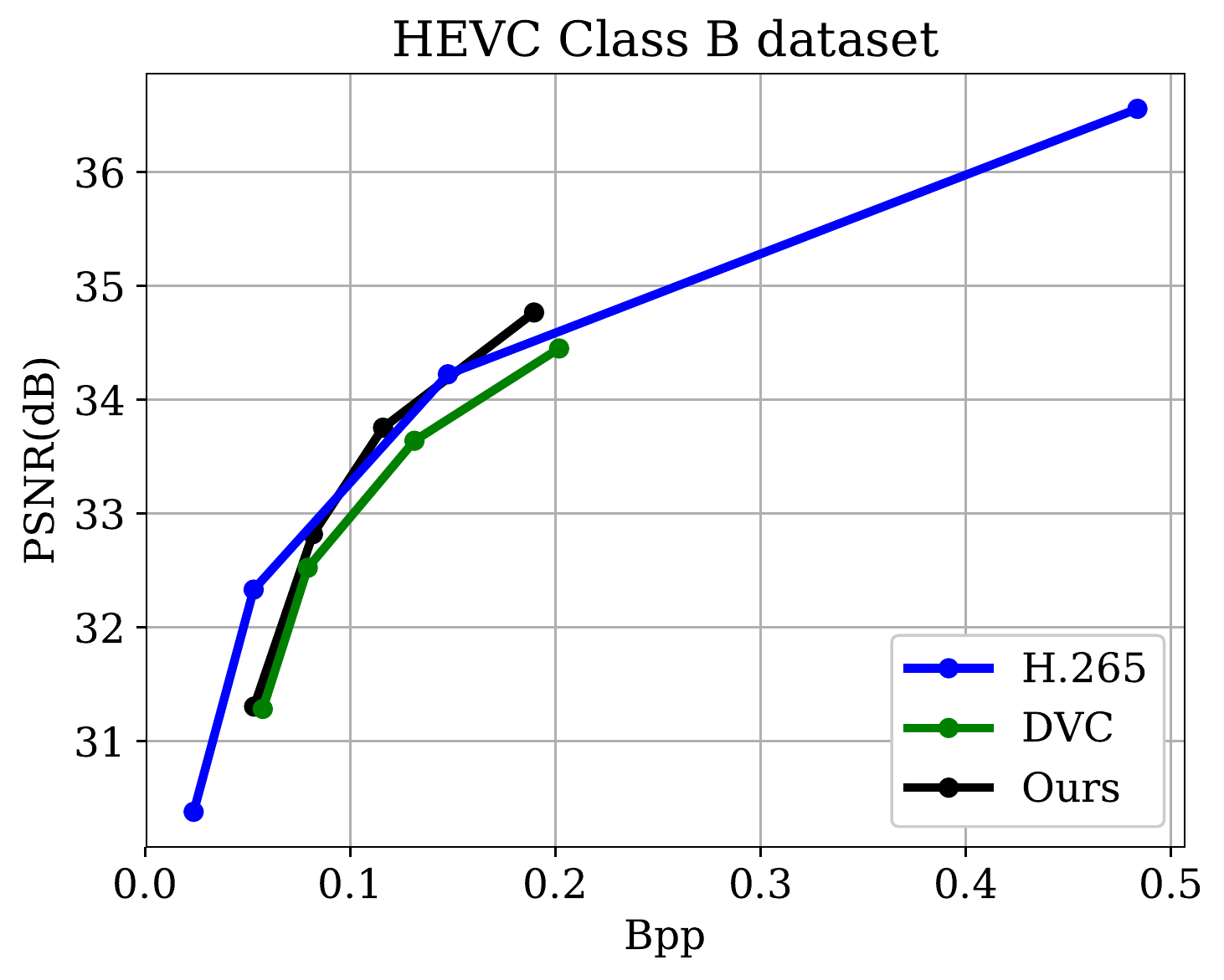}}
  \end{minipage}
    \hspace{\fill}
  \begin{minipage}{0.32\textwidth}
    \centerline{\includegraphics[width=\linewidth]{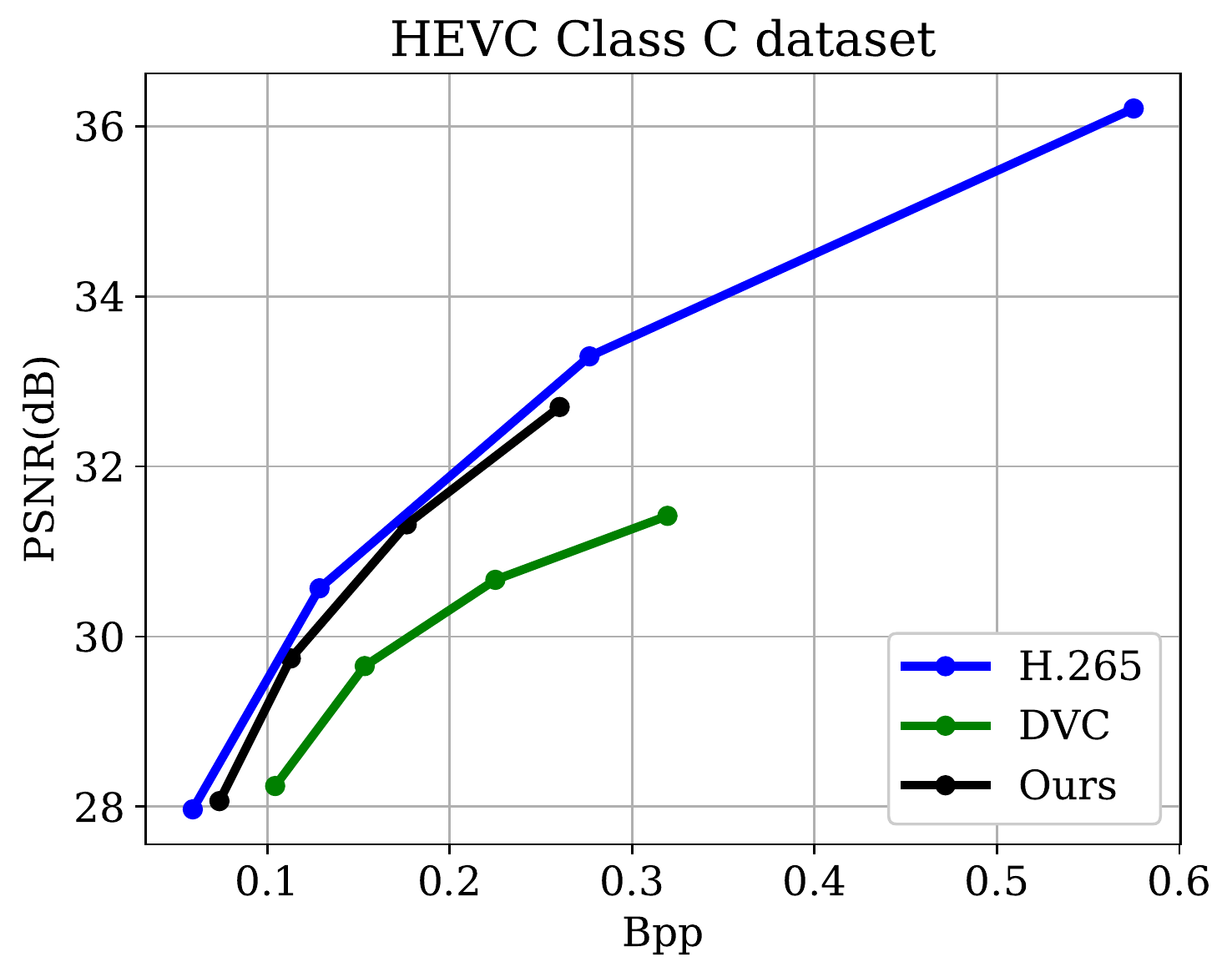}}
  \end{minipage}
      \hspace{\fill}
  \begin{minipage}{0.32\textwidth}
    \centerline{\includegraphics[width=\linewidth]{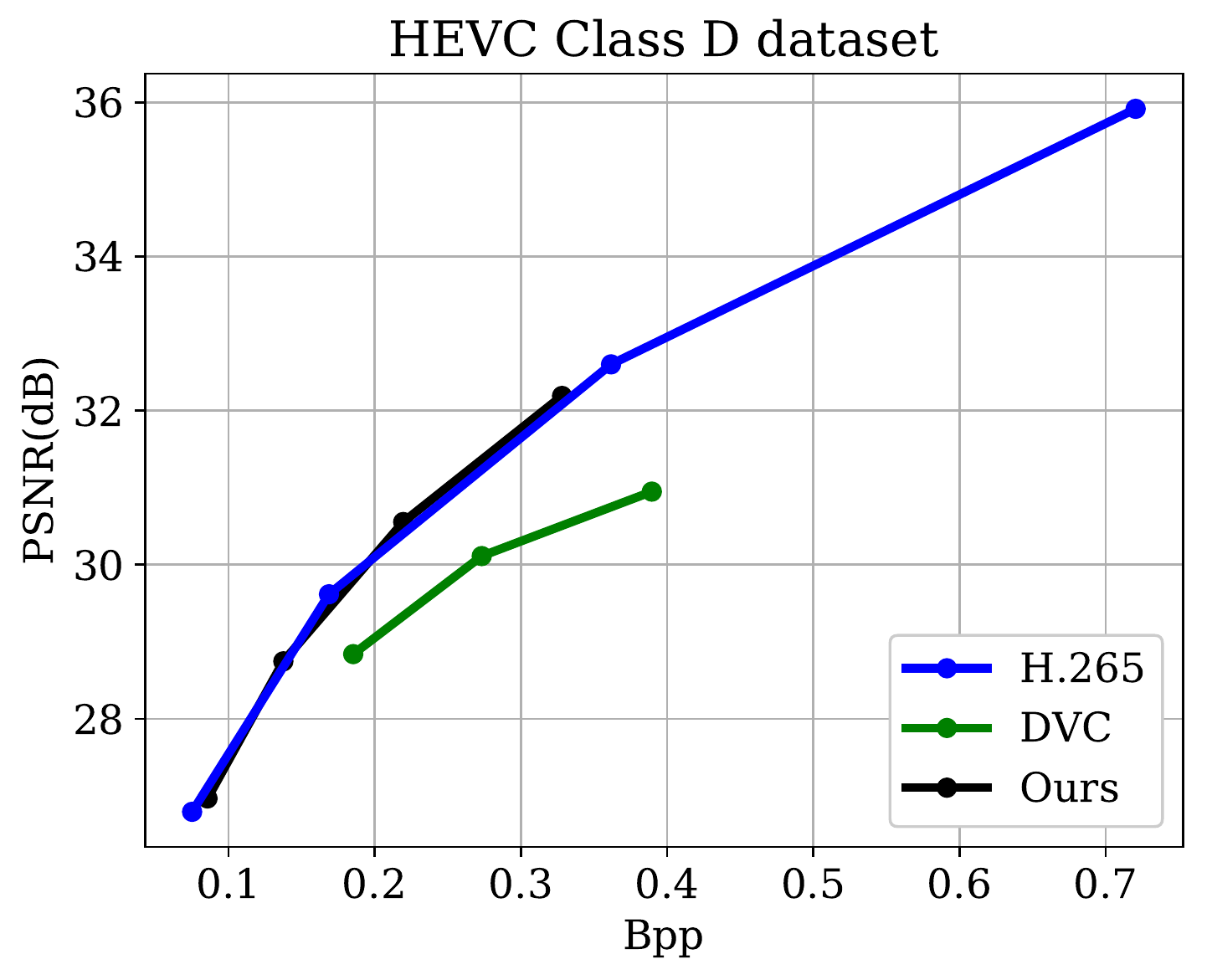}}
  \end{minipage}  
  \caption{Evaluation results for all video frames on the HEVC Class B, Class C and Class D at the default setting.}
  \label{fig:default}
\end{figure*}

For the common testing cases of the traditional video codecs, the GoP size is usually not fixed. To further evaluate the performance of the learning based video codec and the traditional video codec (\textit{e.g.}, H.265), we do not impose any restriction on the GoP size in the codec. 
Specifically, we adopt \textit{veryfast} mode in FFmpeg with the \textit{\textbf{default Setting}}\footnote{ffmpeg -pix\_fmt yuv420p -s WxH -r 50 -i video.yuv  -c:v libx265 -preset veryfast  -tune zerolatency -x265-params ``qp=Q" output.mkv; Q is the quanrization parameter.  W and H are the height and width of the yuv video.}.
We evaluate the compression performance for all the video frames on the HEVC Class B, Class C and Class D datasets. The experimental results are provided in Fig.\ref{fig:default}.

\noindent
\textbf{Baseline Algorithms}
The learning based codecs in Wu \textit{et al.} \cite{Wu_2018_ECCV} and Djelouah \textit{et al.} \cite{DjelouahICCV2019} are based on frame interpolation and designed for B-frame video compression, while the methods in \cite{lu2019dvc,habibian2019video} are for P-frame based video compression. Since the B-frame based compression methods employ two reference frames,  the coding performance is generally better than P-frame based compression method \cite{sullivan2012overview}.
We use the P-frame based compression method DVC \cite{lu2019dvc} as our baseline algorithm and we also demonstrate that the proposed method outperforms all the learning based methods, including the B-frame based compression methods \cite{Wu_2018_ECCV,DjelouahICCV2019}.

\begin{table}[t]
    \caption{The BDBR and BD-PSNR results of different algorithms when compared with H.264. Negative values in BDBR represent the bitrate saving.}
    \centering
    \begin{tabular}{|x{2cm}|x{1cm}|x{1cm}|x{1cm}|x{1cm}|x{1cm}|x{1cm}|}
    \hline
    & \multicolumn{3}{c|}{BDBR(\%)} & \multicolumn{3}{c|}{BD-PSNR(dB)} \\
       \hline  Dataset& H.265 & DVC & Ours & H.265 & DVC & Ours \\    
        \hline Class B &-32.0 &  -27.9 & -41.7 &  0.78 & 0.71 & 1.12\\        
        \hline Class C &-20.8  &-3.5 & -25.9 &  0.91 & 0.13& 1.18\\
        \hline Class D & -12.3  &-6.2  &-25.1   & 0.57  &0.26 &1.25\\
        \hline
    \end{tabular}
    \label{tab:Main_Table}
\end{table}

\noindent
\textbf{Quantitative Evaluation at the \textit{fixed GoP setting}.}
As shown in Fig.\ref{fig:mainresults}, we provide the compression performance of different methods on the HEVC Common Test Sequences. 
When compared with the baseline  DVC \cite{lu2019dvc} algorithm, our proposed method significantly improves compression performance. 
For example,  our proposed method has about 1dB improvement on the HEVC Class C dataset at 0.3bpp. 
It is also observed that the proposed method outperforms the H.264 algorithm and is comparable with H.265 in terms of PSNR.
The BDBR and BD-PSNR results when compared with H.264  are provided in Table \ref{tab:Main_Table}.

We also provide the experimental results when the distortion is evaluated by MS-SSIM. As shown in Fig.\ref{fig:mainresults}, our approach outperforms H.265 in terms of MS-SSIM. One possible explanation is that the traditional codecs \cite{wiegand2003overview,sullivan2012overview} use block based coding scheme, which inevitably generates the block artifacts.

In Fig.\ref{fig:VTL}, Fig.\ref{fig:MCL} and Fig.\ref{fig:UVG},  we evaluate the compression performance on the MCL-JVC, VTL and UVG datasets. 
We compare our proposed method with the recent learning based method \cite{DjelouahICCV2019}, which utilizes B-frame based compression scheme.
As shown in Fig.\ref{fig:VTL}, although we only use one reference frame, the proposed method still achieves better compression performance on the VTL dataset.

In Fig.\ref{fig:UVG}, we compare the proposed method with another state-of-the-art learning based video compression method \cite{habibian2019video} on the UVG dataset.
For a fair comparison with \cite{habibian2019video}, we also use MS-SSIM as the loss function to optimize the network.
The experimental results in Fig.\ref{fig:UVG} demonstrate that the proposed approach outperforms \cite{habibian2019video} by a large margin.

\noindent
\textbf{Quantitative Evaluation at the \textit{default setting}.}
In this section, we also provide comparison results when the traditional codecs use variable GoP size. As shown in Fig.\ref{fig:default}, our method outperforms the previous DVC algorithm \cite{lu2019dvc} by a large margin, especially for the HEVC Class C dataset. A possible explanation is that the error propagation is more severe as the GoP size becomes larger,  which means our proposed scheme will bring more improvements.
Although the proposed method cannot outperform the H.265 at the default setting, the compression performance of these two methods is generally comparable.
Considering that the traditional video codecs exploit other coding techniques, such as multiple reference frames or adaptive quantization parameters, which are not used by current learning based video compression systems, it is possible to further improve the performance of learning based video codec in the future.

\begin{figure*}[!t]
  \begin{minipage}[t]{0.5\textwidth}
    \centerline{\includegraphics[width=\linewidth]{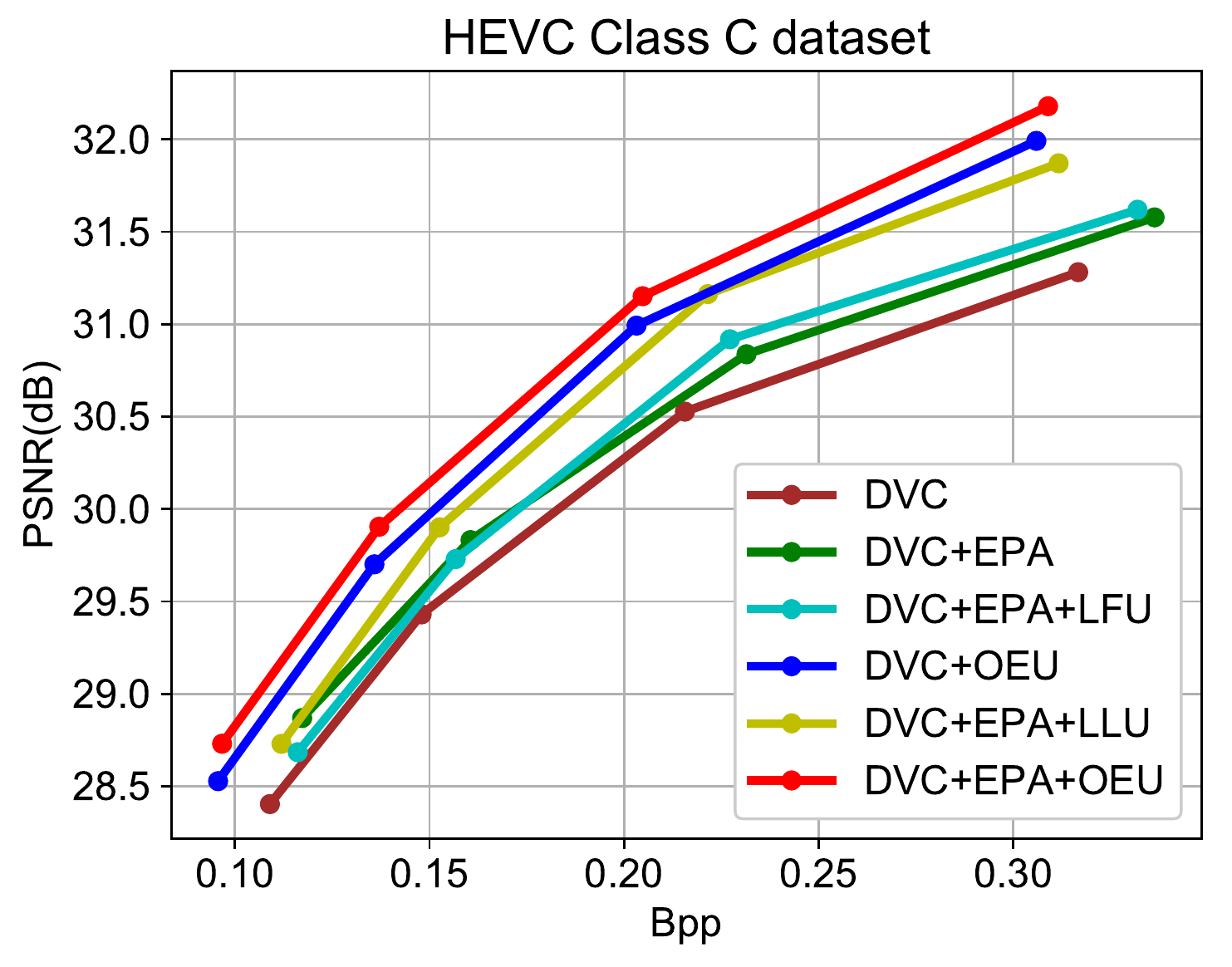}}
    \caption{Ablation study.}
  \label{fig:Ablation}
  \end{minipage}
    \hspace{\fill}
  \begin{minipage}[t]{0.48\textwidth}
    \centerline{\includegraphics[width=\linewidth]{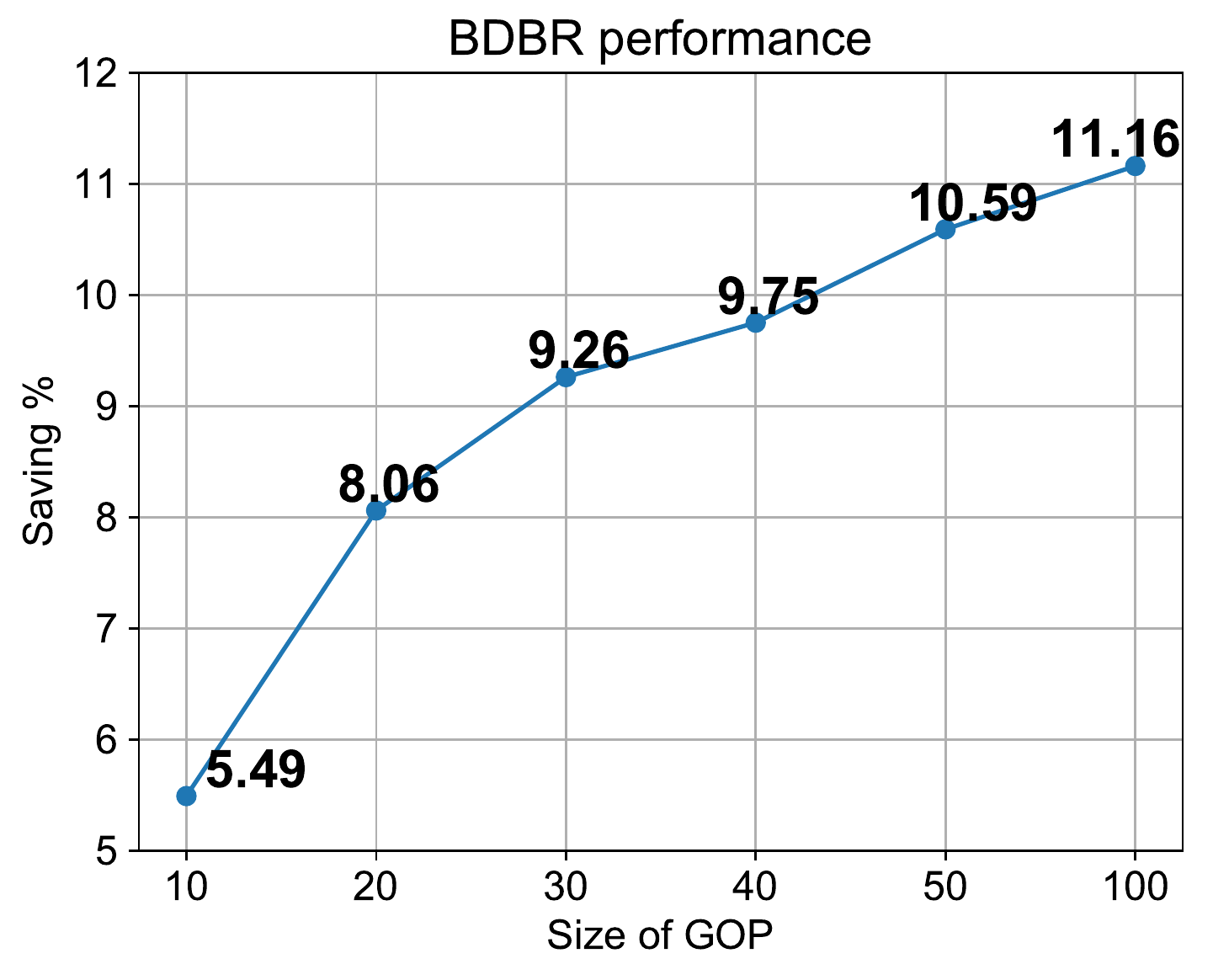}}
    \caption{The bitrate saving when comparing  DVC+EPA with DVC \cite{lu2019dvc} at different GoP sizes.}
  \label{fig:GOP_Ana}
  \end{minipage}
\end{figure*}


\subsection{Ablation Study}

\textbf{The Error Propagation Aware Training Scheme.} 
To demonstrate the effectiveness of our proposed error propagation aware training strategy, we compare the compression performance of different methods in Fig. \ref{fig:Ablation}. 
Specifically, the \textcolor[rgb]{0.43,0.21,0.1}{brown line} represents the DVC algorithm \cite{lu2019dvc}, while the \textcolor[rgb]{0.0,0.5,0.0}{green line} represents the DVC algorithm with the error propagation aware (EPA) training strategy.
It is noticed that the proposed training scheme improves the performance by 0.2dB on the HEVC Class C dataset(GoP=20), which demonstrates that the proposed scheme can alleviate error accumulation by considering temporal neighboring frames in the training stage.


In practical applications, the GoP size for video compression is usually set as 50 or larger to reduce the bandwidth. And error accumulation is more severe as the GoP size increases. 
In Fig.\ref{fig:GOP_Ana}, we investigate the effectiveness of our newly proposed error propagation aware training scheme when the GoP sizes are set as different numbers. We use BDBR \cite{bjontegaard2001calculation} to measure the bitrate saving when compared with the baseline DVC method \cite{lu2019dvc}.
Specifically, the proposed scheme saves 5.49\% bitrate when the GoP size is set to 10 and saves up to 10.59\% bitrate when the GoP size is set to 50. 
The experimental results demonstrate that the proposed method has achieved better compression performance for video sequences with the large GoP size.

\begin{table}[!t]
    \caption{Bitrate saving (BDBR) at different time intervals (\textit{i.e.,} \textit{T} in Eq.\eqref{eq:TAL}).}
    \centering
    \begin{tabular}{|x{1cm}|x{1cm}|x{1cm}|x{1cm}|x{1cm}|x{1cm}|}
        \hline \textit{T} & 2 & 3 & 4 & 5 & 6\\        
        \hline  BDBR & -0.42& -2.12& -3.68& -5.59 & -5.61\\
        \hline
    \end{tabular}
    \label{tab:T_Ana}
\end{table}

To further investigate the proposed error propagation aware training strategy, we provide the compression results when the method is optimized by using different time intervals \textit{T}.
As shown in Table \ref{tab:T_Ana}, the proposed scheme saves more bitrates when $T$ increases. 
For example, the proposed training scheme saves 2.12\% bitrate when setting $T=3$, while the corresponding bitrate saving is 5.59\% when setting $T=5$. 
One explanation is that the objective function can use long-term temporal information when  $T$ increases, which effectively alleviates error accumulation.
In our experiments, we set $T$ to 5 by default.

\noindent
\textbf{The Online Encoder Updating Scheme}
To demonstrate the effectiveness of our proposed online encoder updating~(OEU) scheme in the inference stage, we compare the compression performance of the baseline algorithm with or without using our updating scheme.
In Fig.\ref{fig:Ablation}, the proposed online encoder updating scheme (DVC+OEU, the \textcolor{blue}{blue line}) significantly improves the compression performance by more than 0.5dB. 
Besides, the \textcolor[rgb]{0.7,0.11,0.11}{red line} represents the full model of our proposed method, which achieves the best compression performance by using both the online updating scheme and the error propagation aware training strategy.

In \cite{djelouahcontent}, Campos \textit{et al.} adaptively refined the latent representations of the learning based image codecs for better compression performance. 
Furthermore, we provide the experimental result for the latent features updating (LFU) scheme, where $\hat{m}_t$ and $\hat{y}_t$ are updated and the encoder itself is fixed. 
The corresponding RD curve (DVC+EPA+LFU) is depicted by the \textcolor{cyan}{cyan line}.
Compared with our proposed training scheme (DVC+EPA),  we observe that the performance can be further improved by optimizing the latent representation at a high bitrate.
However, it is obvious that adaptively optimizing the whole encoder (\textcolor[rgb]{0.7,0.11,0.11}{red line} in Fig.~\ref{fig:Ablation}) achieves better performance.
A possible explanation is that updating the encoder provides a larger search range and thus it is more likely to obtain an optimal encoder for the current frame.

Besides, we also provide the compression results when only partial neural networks are updated in the inference stage. Specifically, we use the last layers updating (LLU) scheme, where only the last layers in the residual encoder and motion encoder are updated according to the rate-distortion technique. The experimental results are denoted by the  \textbf{\textcolor[rgb]{0.93,0.86, 0.51}{yellow line}} (DVC+EPA+LLU) in Fig.~\ref{fig:Ablation}. It is observed that the partial updating strategy is also useful for compression. However, the performance is inferior to the proposed approach, where all components in the encoder are updated.

\subsection{Computational Complexity Analysis}

In this paper, we use an adaptive encoder in the inference stage to improve compression performance. 
Since the online rate-distortion optimization scheme is required at the encoder side, it will increase the computational complexity. 
However, it is noticed that the numbers of iterations for different video sequences are different.
For the video sequences with simple motion scenes, such as the HEVC Class B dataset, the encoder learned from the training dataset is already near-optimal and it only requires 3 iterations to obtain the optimal parameters.
For the videos with complex motion scenes, such as the HEVC Class C dataset, more iterations are required to learn the optimal encoder. However, we also obtain a larger improvement($\sim$1dB). 
The encoding speed of H.265(HM) for the HEVC Class C  dataset is about 0.1fps, while our approach is 1.4fps. And the baseline algorithm DVC is 7.1fps.

More importantly, a lot of applications, such as video-on-demand applications, are not sensitive to the computational complexity at the encoder side.
Considering that our approach boosts the compression performance without increasing the decoding time, it is feasible to integrate the proposed techniques with other learning based video codecs, such as \cite{Wu_2018_ECCV}, to further improve the compression performance.

\section{Conclusion}

In this paper, we have proposed a content adaptive and error propagation aware deep video compression method.
Our approach alleviates error accumulation in the training stage and achieves content adaptive coding by using the online encoder updating scheme in the inference stage.
The proposed method is fairly simple yet effective and improves compression performance without increasing the model size or decreasing the decoding speed.
The experimental results show that the compression performance of our proposed method outperforms the state-of-the-art learning based video compression methods.



\clearpage
%
%
\bibliographystyle{splncs}
\bibliography{egbib}

\begin{thebibliography}{10}
\providecommand{\url}[1]{\texttt{#1}}
\providecommand{\urlprefix}{URL }
\providecommand{\doi}[1]{https://doi.org/#1}

\bibitem{BPG}
F. bellard, bpg image format. \url{http://bellard.org/bpg/}, accessed:
  2018-10-30

\bibitem{UVG}
Ultra video group test sequences. \url{http://ultravideo.cs.tut.fi}, accessed:
  2018-10-30

\bibitem{VTL}
Video trace library(vtl) dataset. \url{http://trace.kom.aau.dk/}, accessed:
  2018-10-30

\bibitem{WebP}
Webp. \url{https://developers.google.com/speed/webp/}, accessed: 2018-10-30

\bibitem{agustsson2017soft}
Agustsson, E., Mentzer, F., Tschannen, M., Cavigelli, L., Timofte, R., Benini,
  L., Gool, L.V.: Soft-to-hard vector quantization for end-to-end learning
  compressible representations. In: NIPS. pp. 1141--1151 (2017)

\bibitem{agustsson2018generative}
Agustsson, E., Tschannen, M., Mentzer, F., Timofte, R., Van~Gool, L.:
  Generative adversarial networks for extreme learned image compression. arXiv
  preprint arXiv:1804.02958  (2018)

\bibitem{ahmed1974discrete}
Ahmed, N., Natarajan, T., Rao, K.R.: Discrete cosine transform. IEEE
  transactions on Computers  \textbf{100}(1),  90--93 (1974)

\bibitem{balle2016end}
Ball{\'{e}}, J., Laparra, V., Simoncelli, E.P.: End-to-end optimized image
  compression. In: 5th International Conference on Learning Representations,
  {ICLR} (2017)

\bibitem{balle2018variational}
Ball{\'{e}}, J., Minnen, D., Singh, S., Hwang, S.J., Johnston, N.: Variational
  image compression with a scale hyperprior. In: 6th International Conference
  on Learning Representations, {ICLR} (2018)

\bibitem{bjontegaard2001calculation}
Bjontegaard, G.: Calculation of average psnr differences between rd-curves.
  VCEG-M33  (2001)

\bibitem{chen2018learning}
Chen, Z., He, T., Jin, X., Wu, F.: Learning for video compression. arXiv
  preprint arXiv:1804.09869  (2018)

\bibitem{cheng2019learning}
Cheng, Z., Sun, H., Takeuchi, M., Katto, J.: Learning image and video
  compression through spatial-temporal energy compaction. In: Proceedings of
  the IEEE Conference on Computer Vision and Pattern Recognition, CVPR. pp.
  10071--10080 (2019)

\bibitem{choi2019variable}
Choi, Y., El-Khamy, M., Lee, J.: Variable rate deep image compression with a
  conditional autoencoder. arXiv preprint arXiv:1909.04802  (2019)

\bibitem{DjelouahICCV2019}
Djelouah, A., Campos, J., Schaub-Meyer, S., Schroers, C.: Neural inter-frame
  compression for video coding. In: The IEEE International Conference on
  Computer Vision (ICCV) (October 2019)

\bibitem{djelouahcontent}
Djelouah, J.C.M.S.A., Schroers, C.: Content adaptive optimization for neural
  image compression

\bibitem{habibian2019video}
Habibian, A., van Rozendaal, T., Tomczak, J.M., Cohen, T.S.: Video compression
  with rate-distortion autoencoders. arXiv preprint arXiv:1908.05717  (2019)

\bibitem{kingma2014adam}
Kingma, D.P., Ba, J.: Adam: A method for stochastic optimization. arXiv
  preprint arXiv:1412.6980  (2014)

\bibitem{li2017learning}
Li, M., Zuo, W., Gu, S., Zhao, D., Zhang, D.: Learning convolutional networks
  for content-weighted image compression. In: CVPR (June 2018)

\bibitem{lu2019dvc}
Lu, G., Ouyang, W., Xu, D., Zhang, X., Cai, C., Gao, Z.: {DVC}: An end-to-end
  deep video compression framework. In: Proceedings of the IEEE Conference on
  Computer Vision and Pattern Recognition,{CVPR}. pp. 11006--11015 (2019)

\bibitem{Lu_2018_ECCV}
Lu, G., Ouyang, W., Xu, D., Zhang, X., Gao, Z., Sun, M.T.: Deep kalman
  filtering network for video compression artifact reduction. In: ECCV
  (September 2018)

\bibitem{mentzer2018conditional}
Mentzer, F., Agustsson, E., Tschannen, M., Timofte, R., Van~Gool, L.:
  Conditional probability models for deep image compression. In: CVPR. p.~3.
  No.~2 (2018)

\bibitem{minnen2018joint}
Minnen, D., Ball{\'e}, J., Toderici, G.D.: Joint autoregressive and
  hierarchical priors for learned image compression. In: Advances in Neural
  Information Processing Systems. pp. 10771--10780 (2018)

\bibitem{rippel2017real}
Rippel, O., Bourdev, L.: Real-time adaptive image compression. In: ICML (2017)

\bibitem{rippel2018learned}
Rippel, O., Nair, S., Lew, C., Branson, S., Anderson, A.G., Bourdev, L.:
  Learned video compression. arXiv preprint arXiv:1811.06981  (2018)

\bibitem{schwarz2007overview}
Schwarz, H., Marpe, D., Wiegand, T.: Overview of the scalable video coding
  extension of the h. 264/avc standard. IEEE Transactions on circuits and
  systems for video technology  \textbf{17}(9),  1103--1120 (2007)

\bibitem{shensa1992discrete}
Shensa, M.J.: The discrete wavelet transform: wedding the a trous and mallat
  algorithms. IEEE Transactions on signal processing  \textbf{40}(10),
  2464--2482 (1992)

\bibitem{skodras2001jpeg}
Skodras, A., Christopoulos, C., Ebrahimi, T.: The jpeg 2000 still image
  compression standard. IEEE Signal Processing Magazine  \textbf{18}(5),
  36--58 (2001)

\bibitem{sullivan2012overview}
Sullivan, G.J., Ohm, J.R., Han, W.J., Wiegand, T., et~al.: Overview of the high
  efficiency video coding(hevc) standard. TCSVT  \textbf{22}(12),  1649--1668
  (2012)

\bibitem{theis2017lossy}
Theis, L., Shi, W., Cunningham, A., Husz{\'{a}}r, F.: Lossy image compression
  with compressive autoencoders. In: 5th International Conference on Learning
  Representations, {ICLR} (2017)

\bibitem{toderici2015variable}
Toderici, G., O'Malley, S.M., Hwang, S.J., Vincent, D., Minnen, D., Baluja, S.,
  Covell, M., Sukthankar, R.: Variable rate image compression with recurrent
  neural networks. In: 4th International Conference on Learning
  Representations, {ICLR} (2016)

\bibitem{toderici2017full}
Toderici, G., Vincent, D., Johnston, N., Hwang, S.J., Minnen, D., Shor, J.,
  Covell, M.: Full resolution image compression with recurrent neural networks.
  In: CVPR. pp. 5435--5443 (2017)

\bibitem{tsai2018learning}
Tsai, Y.H., Liu, M.Y., Sun, D., Yang, M.H., Kautz, J.: Learning binary residual
  representations for domain-specific video streaming. In: Thirty-Second AAAI
  Conference on Artificial Intelligence (2018)

\bibitem{wallace1992jpeg}
Wallace, G.K.: The jpeg still picture compression standard. IEEE Transactions
  on Consumer Electronics  \textbf{38}(1),  xviii--xxxiv (1992)

\bibitem{MCL}
Wang, H., Gan, W., Hu, S., Lin, J.Y., Jin, L., Song, L., Wang, P.,
  Katsavounidis, I., Aaron, A., Kuo, C.C.J.: Mcl-jcv: a jnd-based h. 264/avc
  video quality assessment dataset. In: 2016 IEEE International Conference on
  Image Processing (ICIP). pp. 1509--1513. IEEE (2016)

\bibitem{wang2019edvr}
Wang, X., Chan, K.C., Yu, K., Dong, C., Change~Loy, C.: Edvr: Video restoration
  with enhanced deformable convolutional networks. In: Proceedings of the IEEE
  Conference on Computer Vision and Pattern Recognition Workshops. pp.~0--0
  (2019)

\bibitem{wang2003multi}
Wang, Z., Simoncelli, E., Bovik, A., et~al.: Multi-scale structural similarity
  for image quality assessment. In: ASILOMAR CONFERENCE ON SIGNALS SYSTEMS AND
  COMPUTERS. vol.~2, pp. 1398--1402. IEEE; 1998 (2003)

\bibitem{wiegand2003overview}
Wiegand, T., Sullivan, G.J., Bjontegaard, G., Luthra, A.: Overview of the h.
  264/avc video coding standard. TCSVT  \textbf{13}(7),  560--576 (2003)

\bibitem{Wu_2018_ECCV}
Wu, C.Y., Singhal, N., Krahenbuhl, P.: Video compression through image
  interpolation. In: ECCV (September 2018)

\bibitem{xue2017video}
Xue, T., Chen, B., Wu, J., Wei, D., Freeman, W.T.: Video enhancement with
  task-oriented flow. International Journal of Computer Vision, IJCV
  \textbf{127}(8),  1106--1125 (2019)

\end{thebibliography}
\end{document}